\pdfoutput=1
\documentclass[a4paper,10pt,twocolumn]{scrartcl}

\usepackage[latin1]{inputenc}  
\usepackage{amsmath}           
\usepackage{amsfonts}          
\usepackage{amssymb}           
\usepackage{graphicx}          
\usepackage[T1]{fontenc}       
\usepackage{ae}                
\usepackage{typearea}	        
\usepackage{scrpage2}          
\usepackage{textcomp}
\usepackage[margin=10pt,font=small,labelfont=bf]{caption} 

\usepackage[sort&compress,square,comma,authoryear]{natbib}

\usepackage{hyperxmp}
\usepackage[pdfauthor={Thomas Keitzl, Juan-Pedro Mellado, Dirk Notz},pdftitle={Reconciling heat-flux and salt-flux estimates at a melting ice--ocean interface},pdfcreator={LaTeX},pdfcreationdate={D:20160601100000},pdfmoddate={D:20160601100000},pdfkeywords={Free convection, Direct numerical simulation, ice-ocean interaction, Diffusive convection, Double Diffusion, Flux ratio},colorlinks=true,urlcolor=blue,citecolor=black,linkcolor=black,bookmarks=true]{hyperref}
\usepackage{lineno}

\typearea{16}                  
\newcommand{\?}[1]{}
\renewcommand{\!}[1]{}

\begin{document}
\cfoot{\thepage / 17}     	

\twocolumn[{\csname @twocolumnfalse\endcsname                
\titlehead{                                                  
	\begin{tabular*}{\textwidth}[]{@{\extracolsep{\fill}}lr}   
	This is just the preprint of& June 1, 2016\\                  
	\end{tabular*}                                           
	}
\title{Reconciling heat-flux and salt-flux estimates\\at a melting ice--ocean interface}
\author{
  Thomas Keitzl$^1$, Juan-Pedro Mellado$^1$, Dirk Notz$^1$\\ 
  \normalsize{$^1$ Max Planck Institute for Meteorology, Bundesstra\ss{}e 53, 20146 Hamburg, Germany}\\
  \normalsize{\textit{Correspondence to:} T.K. (t@keitzl.com)
}}
\date{}  


\maketitle                                                     
\vspace{-8ex}                                                  
\begin{abstract}  
  The ratio of heat and salt flux is employed in ice--ocean models to represent ice--ocean interactions. 
  In this study, this flux ratio is determined from direct numerical simulations of free convection beneath a melting, horizontal, smooth ice--ocean interface. 
  We find that the flux ratio at the interface is three times as large as previously assessed based on turbulent-flux measurements in the field. 
  As a consequence, interface salinities and melt rates are overestimated by up to 40\% if they are based on the three--equation formulation. 
  We also find that the interface flux ratio depends only very weakly on the far-field conditions of the flow. 
  Lastly, our simulations indicate that estimates of the interface flux ratio based on direct measurements of the turbulent fluxes will be difficult because at the interface the diffusivities alone determine the mixing and the flux ratio varies with depth. 
  As an alternative, we present a consistent evaluation of the flux ratio based on the total heat and salt fluxes across the boundary layer and reconcile the determinations of the ice--ocean interface conditions. 
\\
\\
\end{abstract}
}]

\section{Introduction}
The ice--ocean heat flux is a diffusive flux. 
At the ice--ocean interface, no-slip and no-penetration conditions prohibit any turbulent contribution to the exchange of mass, momentum and energy and only diffusive exchange remains. 
Nonetheless, it is common practice to measure and model the ice--ocean fluxes based on only turbulent fluxes \citep{mcphee_revisiting_2008,sirevaag_turbulent_2009}. 
Here, we present a consistent evaluation of the molecular and turbulent contributions to ice--ocean fluxes and their implication for ice--ocean models. 

Most ice--ocean models conform to the so-called three-equation formulation of the melt rate \citep{holland_modeling_1999}. 
This formulation improves the ice--ocean heat flux significantly over other simplified formulations \citep{schmidt_iceocean_2004}. 
It requires, however, knowledge of the ratio between the fluxes of heat and salt at the interface. 
This flux ratio remains uncertain. 
It has been assessed by various means---from 
  laboratory experiments \citep{martin_experimental_1977} over modeling work \citep{holland_modeling_1999,notz_impact_2003} to field observations \citep{sirevaag_turbulent_2009}. 
  The interface flux ratio, normalised by far-field conditions, has been estimated to range between 20 to 90. 
As demonstrated below, this uncertainty in the flux ratio can cause an uncertainty in modeled heat fluxes of up to 40\%. 
Modelled ice--ocean heat fluxes are part of global circulation models and local ice--ocean interaction models, both which desire the fluxes to be physically correct. 

In this work, we strive to settle the debate on the one flux ratio. 
First, we introduce the formalism that is generally employed to determine the interface conditions and the heat flux from the flux ratio. 
Along we present its historic context (second section). 
Then, we explain the details of our setup and method: a direct numerical simulation of free convection beneath a melting, horizontal, smooth ice--ocean interface in the semiconvective regime (third and fourth section). 
We employ the simulation in the following to separately investigate the molecular and the turbulent contribution to the vertical structure of the flux ratio (fifth section). 
This investigation shows that all of the former assessed flux ratios may be reasonable, and we explain the rationale behind this. 
Finally, we discuss the interface value of the flux ratio which determines the ice--ocean interface conditions and hence the melt rates (sixth section). 
We find that the interface flux ratio is almost independent of far-field conditions and three times as large as the value previously estimated from field measurements. 
As a result, heat-flux parameterisations based on the three-equation formulation can lead to melt rates over-estimated by up to 40\%.

\section{Formalism}
The ratio between heat flux and salt flux is a relevant quantity of the ice--ocean formalism, 
  because it determines the interface values of temperature and salinity, $T_\textrm{i}$ and $S_\textrm{i}$. 
This is well illustrated by 
  the boundary conditions of an ice--ocean interface, 
\begin{linenomath}
\begin{subequations}
\label{eq_bc}
\begin{align} 
&F_\textrm{h,i} = \rho_\textrm{ice} w_0 L \label{eq_bc2}\\	
&F_\textrm{s,i} = \rho_\textrm{ice} w_0 S_\textrm{i} \label{eq_bc3} \\
&T_\textrm{i} = - m S_\textrm{i}\label{eq_bc1} 	 \textrm{:}      
\end{align}
\end{subequations}
\end{linenomath}
Because the ice of density $\rho_\textrm{ice}$ can only dissolve and melt at one particular rate, $w_0$, 
the interface salinity must depend on the heat flux, $F_\textrm{h}$, and the salt flux, $F_\textrm{s}$, according to Eq. (\ref{eq_bc2}) and Eq. (\ref{eq_bc3}) as $S_\textrm{i} = L F_\textrm{s,i} / F_\textrm{h,i}$. 
The index $i$ denotes interface values and 
$L$ is the latent heat of fusion. 
From the interface salinity, the interface temperature follows with Eq. (\ref{eq_bc1}), where $m$ describes the freezing point relation. 
The challenging part about determining the interface conditions, $T_\textrm{i}$ and $S_\textrm{i}$, is a proper representation of the fluxes, $F_\textrm{s}$ and $F_\textrm{h}$, or their ratio.

\cite{josberger_sea_1983} applies both a bulk description and a detailed description based on insights by \cite{mcphee_analytic_1981} to represent these fluxes. 
From both descriptions he finds that the interface conditions depend on the far-field conditions and on only one property relating to the flow beneath the ice: 
the non-dimensional ratio of the heat flux to the salt flux at the interface, 
\begin{linenomath}
\begin{subequations}
\label{eq_gamma}
\begin{align}
\gamma_\textrm{i}=\frac{F_\textrm{h,i}/ \left[ c_p \left(T_\infty - T_\textrm{i}\right)\right]}{F_\textrm{s,i}/\left(S_\infty - S_\textrm{i}\right)} \textrm{,}
\end{align}
\end{subequations}
\end{linenomath}
where $c_p$ is the specific heat capacity of water. 
This ratio describes how effectively turbulence mixes heat compared to salt near the interface. 
We reproduce the resulting interface salinity from Josberger's three--equation formulation from Eq. (\ref{eq_bc}) with the commonly used bulk flux parameterisations,
\begin{linenomath}
\begin{subequations}
\label{eq_flux_bulk}
\begin{align}
&F_\textrm{h,i} = \rho_\textrm{water} c_p \alpha_\textrm{h} u_{*0} \left(T_\infty - T_\textrm{i}\right) \label{eq_alpha_h} \textrm{and}\\
&F_\textrm{s,i} = \rho_\textrm{water} ~~~~ \alpha_\textrm{s} u_{*0} \left(S_\infty - S_\textrm{i}\right) \label{eq_alpha_s} \textrm{,}
\end{align}
\end{subequations}
\end{linenomath}
which are based on the bulk heat exchange coefficient, $\alpha_\textrm{h}$, the bulk salt exchange coefficient, $\alpha_\textrm{s}$, and the friction velocity $u_{*0}$ \citep{notz_impact_2003}. 
By substituting Eq. (\ref{eq_flux_bulk}) into Eq. (\ref{eq_bc}), one obtains 
\begin{linenomath}
\begin{subequations}
\label{eq_salinity_interfacial_quadratic}
\begin{align}
&m S_\textrm{i}^2 + \left(T_\infty + \Delta T_\gamma \right) S_\textrm{i} - \Delta T_\gamma S_\infty = 0 \label{eq_salinity_interfacial_quadratic_a}\\ 
&\Delta T_\gamma = L c_p^{-1} \gamma_\textrm{bulk}^{-1} \textrm{.}
\end{align}
\end{subequations}
\end{linenomath}
As found by \cite{josberger_sea_1983}, $\gamma_\textrm{bulk} = \alpha_\textrm{h}/\alpha_\textrm{s}$ is the one flow property that determines the interface conditions from the far-field temperature, $T_\infty$, and far-field salinity, $S_\infty$. 

\cite{holland_modeling_1999} find from different modeling approaches that $\gamma_\textrm{bulk} > 1$ and almost independent of the far-field mean shear velocity. 
The different model approaches that they employ yield flux ratios between 25 and 200 according to their Figure 4. 

\cite{notz_impact_2003} take on the determination of $\gamma_\textrm{bulk}$ from field measurements. 
By modeling observations of false-bottom persistence and migration under sea ice, they indirectly show that $\gamma_\textrm{bulk}$ needs to be substantially different from unity. 
They adapt results of laboratory studies of fluid heat and mass exchange across hydraulically rough surfaces to describe the dependence of the bulk exchange coefficients on the molecular diffusivities \citep{mcphee_dynamics_1987}. 
Within the range they estimate for $\gamma_\textrm{bulk}$ [$35 < \gamma_\textrm{bulk} < 70$ \citep{yaglom_heat_1974,owen_heat_1963}], they find that a value on the higher end of the range fits their data better. 

In sea-ice literature, the bulk exchange coefficients are generally referred to as interface exchange coefficients or turbulent exchange coefficients (not to be confused with eddy diffusivity, which is commonly referred to as turbulent exchange coefficient). 
The interface fluxes of heat and salt are further approximated by the turbulent fluxes in a certain distance from the interface. 
In these lines, \cite{sirevaag_turbulent_2009} follows up on the efforts of \cite{notz_impact_2003} to determine a turbulent flux ratio, $\gamma_\textrm{turb}$, from direct field measurements at a certain distance from the interface. 
From turbulent-instrument-cluster measurements 1~m beneath the ice--ocean interface in the area of Whaler's Bay, he determines average temperature, salinity, friction velocity and turbulent fluxes of heat and salt. 
From his results 
  ($\alpha_\textrm{h} = 1.31 \times 10^{-2}$, $\alpha_\textrm{s} = 4.0  \times 10^{-4}$) he estimates $\gamma_\textrm{turb}\approx33$ (or $\gamma_\textrm{turb}\approx23$ if only data with small mean-temperature changes is accounted for), a value that lies on the opposing end of the range and value given by \cite{notz_impact_2003}. 

As opposed to the above mentioned application of the bulk parameterisation [see Eq. (\ref{eq_flux_bulk})] and turbulent-flux measurements, \cite{gade_when_1993} applies the diffusive flux definitions,
\begin{linenomath}
\begin{subequations}
\label{eq_flux_interface}
\begin{align}
&F_\textrm{h,i} = \rho_\textrm{water} c_p \kappa_\textrm{t}\left. \partial_z T \right|_{z_\textrm{i}} \label{eq_flux_interface_T} \textrm{, and}\\
&F_\textrm{s,i} = \rho_\textrm{water} ~~~~ \kappa_\textrm{s} \left. \partial_z S \right|_{z_\textrm{i}} \label{eq_flux_interface_S} \textrm{,}
\end{align}
\end{subequations}
\end{linenomath}
to determine the interface conditions.
In line with his procedure, one introduces the gradient thickness of temperature, $\delta_\textrm{t}= \left(T_\infty - T_\textrm{i}\right) / \left. \partial_z T \right|_{z_\textrm{i}}$, and salinity analogously, 
and ends up with Eq. (\ref{eq_salinity_interfacial_quadratic}), but with $\Delta T_\gamma = L c_p^{-1} \gamma_\textrm{mol}^{-1}$, where $\gamma_\textrm{mol} = \textrm{Le}~\delta_\textrm{s}/\delta_\textrm{t}$ is a molecular flux ratio, $\textrm{Le}=\kappa_\textrm{t}/\kappa_\textrm{s}$ is the Lewis number, and $z_\textrm{i}$ is the position of the ice--ocean interface.  
He finds a substantially different flux ratio. 
Based on the experimental work by \cite{martin_experimental_1977}, \cite{gade_when_1993} determines the boundary thickness ratio, 
\begin{linenomath}
\begin{align}
R=\delta_\textrm{t}/\delta_\textrm{s} \textrm{,} \label{eq_bl_thickness_ratio}
\end{align}
\end{linenomath}
to $2.3$ which yields $\gamma_\textrm{mol} = 88.9$, for $\kappa_\textrm{t}=1.39 \times 10^{-7}~\textrm{m}^2~\textrm{s}$ and $\kappa_\textrm{s}=6.8 \times 10^{-10}~\textrm{m}^2~\textrm{s}$. 
We stress the difference between the ratio of exchange coefficients that have been determined so far: $\gamma_\textrm{mol} = 88.9$ and $\gamma_\textrm{turb}\approx33$ (from above). 
  
The uncertainty in the flux ratio leads to considerable uncertainty in the determination of the interface conditions. 
\begin{figure*}[tbp]
\begin{center}
\includegraphics[width=.7\linewidth]{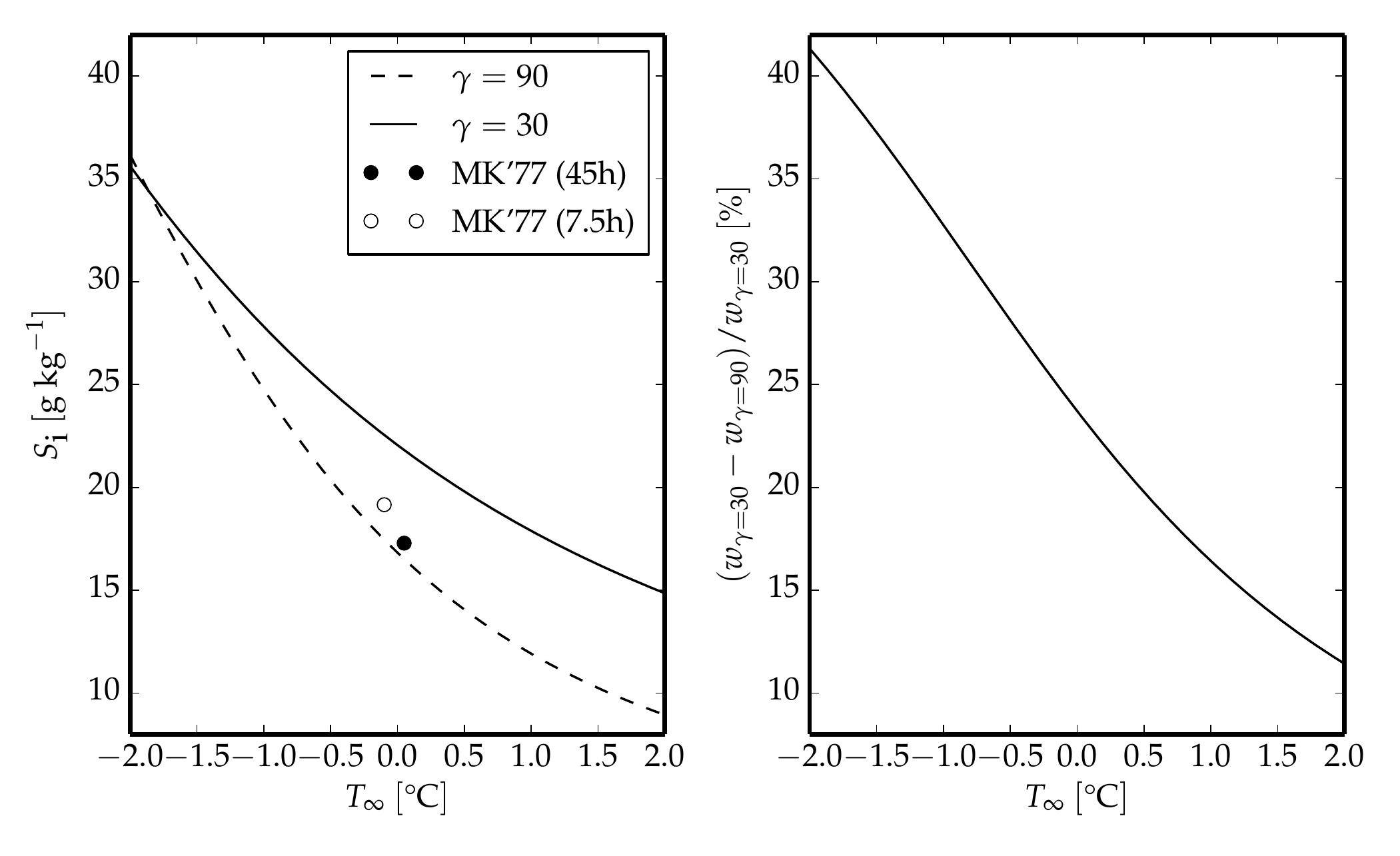}
\caption{\small{
a) Theoretical interfacial salinity, $S_\textrm{i}$, for $S_\infty=34~\textrm{g}~\textrm{kg}^{-1}$ and varying $T_\infty$ with $\gamma_\textrm{i}=30$ or $\gamma_\textrm{i}=90$. 
   Measured interfacial salinites 7.5~h (empty dot) and 47~h (filled dot) after laboratory-experiment start \citep{martin_experimental_1977}. 
b) Relative difference in melt rate due to difference in interface salinity as seen in (a). 
   The melt rate, $w_0$, is given from Eq. (\ref{eq_bc2}), Eq. (\ref{eq_alpha_h}), with $\alpha_\textrm{h}=\textrm{const.}$, $u_{*0}=\textrm{const.}$, and $T_\textrm{i}=T_\textrm{i}(\gamma)$ according to Eq. (\ref{eq_salinity_interfacial_quadratic}).
}}
\label{fig_dev} 
\end{center}
\end{figure*}
Whether $\gamma_\textrm{i}$ equals $\gamma_\textrm{turb}$ ($\approx 30$) \citep{sirevaag_turbulent_2009} or $\gamma_\textrm{mol}$ ($\approx 90$) \citep{gade_when_1993} leads to a different determination of the interface salinity according to Eq. (\ref{eq_salinity_interfacial_quadratic}) (see Figure \ref{fig_dev}a). 
The temperature difference between far field and interface that follows from the different interface salinities is significant. 
That implies an overestimation of the melt rate by up to 40\% with the value of $\gamma_\textrm{turb}\approx30$ as compared to $\gamma_\textrm{mol}\approx90$ (see Figure \ref{fig_dev}b).

\section{Setup}
A mass of solid ice rests on top of a body of sea water of fixed uniform temperature $T=T_\infty$ and salinity $S=S_\infty$. 
Ice and water form a horizontal interface. 
The ice is isothermal at the freezing temperature of the interface, isohaline at the interface salinity and has a smooth surface. 
The ice imposes the boundary conditions given in Eq. (\ref{eq_bc}) on the temperature and salinity fields and no-slip, no-penetration boundary conditions on the flow field. 
We consider the ice--ocean interface together with the sea-water body as our system of interest. 

\subsection{The Evolution of the System}
This system is purely buoyancy driven and evolves in space and time according to the evolution equations of mass, momentum, internal-energy and solute. 
With the velocity field $\textbf{v}\left(\textbf{x},t\right)$, 
    the temperature field $T\left(\textbf{x},t\right)$, 
    the salinity field $S\left(\textbf{x},t\right)$, 
    the spatial coordinate $\textbf{x}=x_1\textbf{e}_1 + x_2 \textbf{e}_2 - z \textbf{e}_3$, 
    with $\textbf{e}_i=\varepsilon_{ijk}\textbf{e}_j \textbf{e}_k$, 
    and with time $t$, 
these evolution equations are 
\begin{linenomath}
\begin{subequations}
\label{eq_evolve}
\begin{align}
  &\partial_j \upsilon_j =0 \label{eq_mass} \textrm{,} \\
  &\partial_t \upsilon_i = -\upsilon_j \partial_j \upsilon_i + \nu \partial_j^2 \upsilon_i -\partial_i p + b(S,T)~\delta_{i3} \label{eq_mom} \textrm{,}\\
  &\partial_t T = -\upsilon_j \partial_j T + \kappa_\textrm{t}\partial_j^2 T \label{eq_energy} \textrm{,} \\
  &\partial_t S = -\upsilon_j \partial_j S + \kappa_\textrm{s} \partial_j^2 S \label{eq_energy_pot} \textrm{.}
\end{align}
\end{subequations}
\end{linenomath}
The equations are given in the Boussinesq approximation. 
$\nu$ is the kinematic viscosity, $\kappa_\textrm{t}$ the thermal diffusivity, $\kappa_\textrm{s}$ the diffusivity of salinity, $p$ the modified kinematic pressure, $\partial_t$ the temporal derivative and $\partial_i$ is the spatial derivative in direction of $\textbf{e}_i$. 

The buoyancy, $b$, depends on both the evolution of temperature and salinity. 
We follow previous numerical work on different double-diffusive systems of \cite{nagashima_three_1997, gargett_direct_2003, kimura_direct_2007, zweigle_direkte_2011, carpenter_simulations_2012} and approximate the buoyancy to first order by 
\begin{linenomath}
\begin{align}
&b(S,T) = \frac{g}{\rho_\textrm{water}} \left[\beta \left(S_\infty - S\right) - \alpha \left(T_\infty - T\right) \right] \textrm{,}
\label{eq_b_bilinear}
\end{align}
\end{linenomath}
$g$ is earth's gravitational acceleration, $\alpha$ is the thermal expansion coefficient, $\beta$ is the haline contraction coefficient, and $\rho_\textrm{water}$ is the water density, and the subscript $_\infty$ denotes the values far away from the interface---in the far field. 
The salinity component of buoyancy stabilises the water column for values $S<S_\infty$, the temperature component destabilises the water column for values $T<T_\infty$. 
From the interplay of both salinity and temperature at different diffusivities follows a buoyancy-reversal instability that forces the system. 

For sufficiently low viscosity the system becomes turbulent, decorrelates from its initial state, and solely depends on the set of control parameters 
$\{\nu,~\kappa_\textrm{t},~\kappa_\textrm{s},~g~\rho_\infty^{-1}~\alpha~\left(T_\infty-T_\textrm{i}\right),~g~\rho_\infty^{-1}~\beta~\left(S_\infty-S_\textrm{i}\right)\}$. 
Dimensional analysis provides the set of non-dimensional, independent control parameters $\{\textrm{Pr},~\textrm{Le},~R_\rho^s\}$, with Prandtl number $\textrm{Pr}=\nu/\kappa_\textrm{t}$, Lewis number $\textrm{Le}=\kappa_\textrm{t}/\kappa_\textrm{s}$, and density ratio 
\begin{linenomath}
\begin{align}
R_\rho^s= \frac{\beta \left(S_\infty-S_\textrm{i}\right)}{\alpha \left(T_\infty-T_\textrm{i}\right)}
\label{eq_density_ratio}
\end{align}
\end{linenomath}
\citep{turner_double-diffusive_1974}. 
For any given fluid of fixed $\textrm{Pr}$ and fixed $\textrm{Le}$, any flow property does only depend on the governing parameter $R_\rho^s$. 
$R_\rho^s$ quantifies the stabilising effect of the salinity component compared to the effect of the destabilising temperature component. 
Eq. (\ref{eq_b_bilinear}) is then written as 
\begin{linenomath}
\begin{align}
\label{eq_b_bilinear_nd}
&\frac{b}{\left| b_m \right|} = R_\rho^s~\sigma  - \theta \textrm{,}
\end{align}
\end{linenomath}
with the minimum buoyancy $b_m=b(S_\infty,T_\textrm{i})$, the normalised salinity $\sigma$, and the normalised temperature $\theta$, 
\begin{linenomath}
\begin{subequations}
\label{eq_sbm_scalars_nd}
\begin{align}
  &\theta = \frac{T_\infty-T}{T_\infty-T_\textrm{i}} \textrm{,}\\
  &\sigma = \frac{S_\infty-S}{S_\infty-S_\textrm{i}} \textrm{.}
\end{align}
\end{subequations}
\end{linenomath}
We assess the range of validity of Eq. (\ref{eq_b_bilinear_nd}) from a parametric $\left(S,~T\right)$ plot that we compare to the proper formulation by \cite{sharqawy_thermophysical_2010} (not shown).  
For $R_\rho^s > 5$, Eq. (\ref{eq_b_bilinear_nd}) becomes increasingly good an approximation with absolute buoyancy deviations less then $0.20~\left|b_m\right|$. 

The fully developed turbulent system is statistically homogeneous in horizontal directions. 
We denote horizontally averaged quantities by $\left\langle\cdotp\right\rangle$ and fluctuations around that mean by $\cdotp^\prime$. 
Horizontally averaged statistics only depend on $\{R_\rho^s;~z,~t\}$ with $z=-~\vec{x} \cdot \hat{e}_3$ and the origin of $\vec{x}$ chosen such that $z$ gives the distance from the interface. 

The flow develops freely into the far field (see Figure \ref{fig_vis}). 
It does not feel any solid boundary but the ice--water interface. 

\subsection{The Boundary Conditions}
The boundary conditions, Eq. (\ref{eq_bc}), are Robin boundary conditions.
The interface temperature and the interface salinity, $T_\textrm{i}=-m~S_\textrm{i}$ and $S_\textrm{i}=L~F_\textrm{s,i}/F_\textrm{h,i}$, 
depend on fluxes, $F_\textrm{s}$ and $F_\textrm{h}$, at the interface. 
The fluxes evolve with the flow and so do the boundary conditions. 
Notwithstanding the boundary conditions encountered in nature, we simplify the system in two respects. 

First, we apply homogeneous and steady Dirichlet boundary conditions at the top boundary of the scalar fields. 
Therefore, the system does only reflect a natural evolution of the flow once the interface flux ratio has reached an equilibrium, 
when the interface temperature and interface salinity are fixed and do no longer evolve with the flow. 
The tendency of a similar system to relax towards a preferred interface flux ratio has been observed before by \cite{carpenter_simulations_2012}. 
In the supplementary material, we show that the flux ratio does also relax towards an equilibrium in our simulations. 

Second, we do not incorporate melt-water formation. 
According to \cite{keitzl_impact_2016} melt-water formation only influences the flow structure when the Richardson number approaches one. 
This Richardson number describes the importance of the stable stratification next to the interface compared to the strength of the buoyancy reversal that drives the convection. 
The Richardson number relates to the density ratio $R_\rho^s$ of the setup described in the present paper. 
\cite{keitzl_impact_2016} suggest to define a Richardson number based on the minimum buoyancy, $b_m=b(S_\infty,T_\textrm{i})$. 
With Eq. (\ref{eq_b_bilinear_nd}), $b_m$ and $b(z_\textrm{i})=b(S_\textrm{i},T_\textrm{i})$, one obtains
\begin{linenomath}
\begin{align}
\frac{b(z_\textrm{i})}{\left| b_m \right|} = &R_\rho^s - 1  \textrm{,}
\label{eq_richardson}
\end{align}
\end{linenomath}
which resembles their definition of the reference Richardson number, $\textrm{Ri}_0$. 
One can hence expect that $R_\rho^s$ similarly describes the importance of the stable stratification next to the interface compared to the strength of the buoyancy inversion. 
For their free-convection system, \cite{keitzl_impact_2016} find that melt-water formation can be neglected as long as the buoyancy inversion cannot compete with the stable stratification next to the ice, that is as long as $R_\rho^s \gg 1$. 
We restrict our investigations to such systems.

\section{Direct Numerical Simulation}
\begin{center}
\begin{figure*}[ht]
\begin{center}
\includegraphics[width=1.\linewidth]{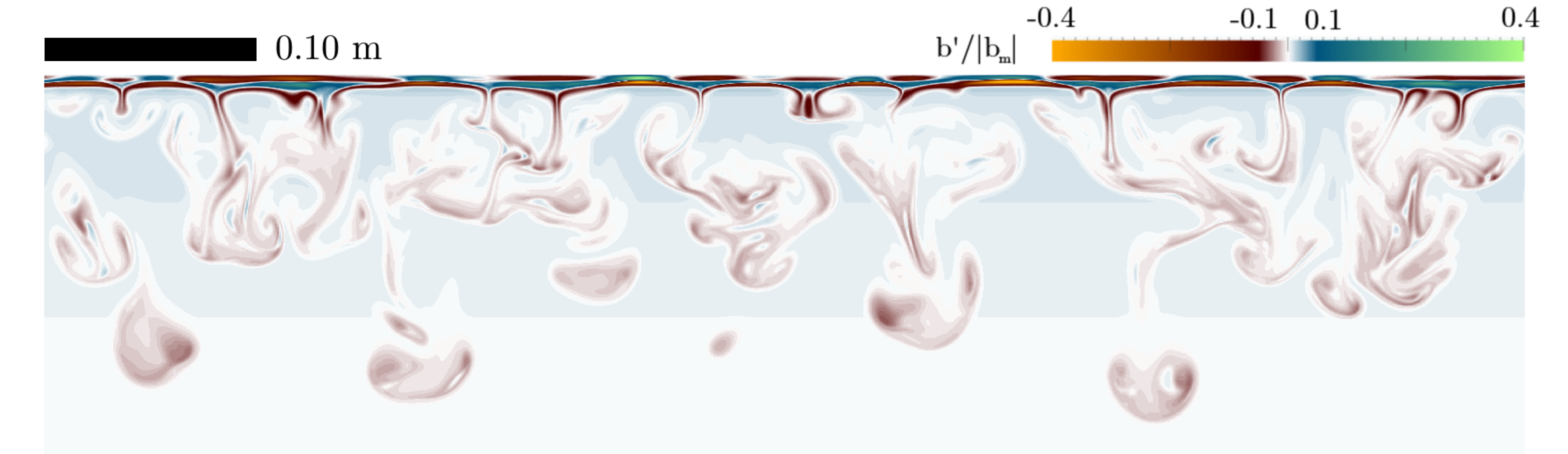}
\caption{\small{
Simulated buoyancy fluctuations at final simulation time for $\textrm{Pr=10}$,~$\textrm{Le=4}$,~and $R_\rho^s=6$ (see Table \ref{tab_simulations_sbm}).
}}
\label{fig_vis} 
\end{center}
\end{figure*}
\end{center}
We integrate Eqs. (\ref{eq_evolve}) using a high-order finite-difference method on a collocated, structured grid. 
We approximate the integration by a fourth-order Runge--Kutta scheme and the spatial derivatives by sixth-order spectral-like finite differences \citep{williamson_low-storage_1980,lele_compact_1992}. 
After every integration step, a pressure solver ensures fulfillment of the solenoidal constraint. 
For this we use a Fourier decomposition along periodic horizontal coordinates and a factorisation of the resulting second-order equations in the vertical coordinate \citep{mellado_factorization_2012}. 

The calculations are performed on a grid of 1152 grid points in the vertical direction and 2560 grid points in both horizontal directions. 
The grid spacing is uniform in the horizontal directions $\textbf{e}_1$, $\textbf{e}_2$ and in most of the vertical direction $\textbf{e}_3$. 
The resolution in $\textbf{e}_3$ close to the interface, however, is increased 
  because the main mean-temperature and mean-salinity variation all over the domain occurs close to the interface. 
These variations potentially entail the main mean-buoyancy change, a change in the forcing of the system from a positive to the global-extreme negative value and back to almost zero. 
To fully cover this buoyancy variation, we increase the resolution next to the interface by a factor of two and a half. 
The regions of uniform and adjusted resolution in $\textbf{e}_3$ are gradually matched by hyperbolic tangents. 
Finally, the grid in $\textbf{e}_3$ far from the interface is coarsened to save computing time. 
This part of the domain serves to diminish the influence of the computational boundary on the flow. 
This grid spacing, $\Delta x_3$, holds $\Delta x_3 / \eta_B < 2.0$ at all times where $\eta_B$ is the Batchelor scale. 

The boundary conditions in the velocity field are no-slip and no-penetration at the interface, and free-slip and no-penetration in the far field. 
The boundary conditions in the temperature and salinity field are Dirichlet at the interface and Neumann in the far field. 
The initial conditions are an error-function profile in the temperature and salinity fields and zero in the velocity field. 
The error-function profiles are described by their gradient thicknesses, $\delta_\textrm{s}$ and $\delta_\textrm{t}$, at initial time. 
The profiles of temperature, salinity and flow fields are perturbed by broadband fluctuations to accelerate the transition to turbulence. 
The initial boundary thickness ratio, $R_\textrm{init}=R(t=0)$, remains the most influential parameter of the initial conditions.
A value around 1.5 or below forces a diffusive evolution in the beginning. 
A larger value forces the system with a buoyancy-reversal instability due to the opposing forcing mechanisms of temperature and salinity. 

We are particularly interested in simulations of $\textrm{Pr}$ between 10--13.8 and $\textrm{Le}$ between 176--204 that resemble cold ocean-like fluids \citep{steele_role_1989,notz_impact_2003,schmidt_iceocean_2004, sharqawy_thermophysical_2010}. 
The available computational resources, however, constrain our investigations to $\textrm{Pr}\times\textrm{Le}=40$ for which turbulence is still fully resolved on diffusive scales. 
For our main simulations we stick to a water-like fluid of $\textrm{Pr}=10$ but of limited $\textrm{Le}=4$. 
The main simulation runs explore the influence of varying $R_\rho^s\in\{6,11,21\}$ (see Table \ref{tab_simulations_sbm}). 
The estimated final boundary-layer height of the simulated systems, $z_\textrm{est}$, is about 0.2~m (see Table \ref{tab_simulations_sbm}).
The simulations reach Reynolds numbers $w_* z_\textrm{est} \nu^{-1}$ and $e^2 {\left(\varepsilon \nu\right)}^{-1}$ of up to 350 and 25, respectively, with the turbulent kinetic energy $e$, the viscous dissipation rate $\varepsilon$, the convective velocity scale, $w_*$ and the viscosity $\nu$. 

To circumvent the computational constraints of three-dimensional simulations, we further conduct two-dimensional simulations. 
The two-dimensional simulations contribute additional evidence in the region of the parameter space accessible to the three-dimensional simulations, and allow us to extrapolate further into the parameter space: 
Simulations of $\textrm{Pr}=1$ approach the behaviour of varying $\textrm{Le}$ up to $\textrm{Le}=160$. 
Simulations of both $\textrm{Pr}=6$ and $\textrm{Pr}=10$ approach the behaviour of varying $\textrm{Le}$ up to $\textrm{Le}\approx20$. 
\begin{center}
\begin{table*}[htbp]     
\begin{tabular*}{\linewidth}{@{\extracolsep{\fill}}cccccccccccccc}
\hline
\hline
\rule[-6pt]{0pt}{23pt}	&	$R_\rho^s$	&$\textrm{Pr}$	&$\textrm{Le}$	&&$T_\infty$~[\textcelsius]	&$\left| b_m \right|$ [m s$^{-2}$]	&	$z_\textrm{est}$ [m]	&$\frac{w_* z_\textrm{est}}{\nu}$	&	$\frac{k^2}{\left(\varepsilon \nu\right)}$	&$R_\textrm{final}$	& \\
\hline
\rule[-6pt]{0pt}{21pt}	&	~6	&	10	&	~4		&&~24.0			& $4.51\times10^{-2}$			&	0.17			&	348				&	25			&	 1.67			&	\\
\rule[-6pt]{0pt}{21pt}	&	11	&	10	&	~4		&&~16.1			& $2.45\times10^{-2}$			&	0.17			&	245				&	18			&	 1.80			&	\\
\rule[-6pt]{0pt}{21pt}	&	21	&	10	&	~4		&&~10.4			& $1.27\times10^{-2}$			&	0.19			&	192				&	13			&	 1.89			&	\\
\hline
\hline
\end{tabular*}  
\caption{Properties of the numerical simulations of the ice--ocean system. 
  The set of the first three columns uniquely defines the system. 
  They are the density ratio, $R_\rho^s$, the Prandtl number $\textrm{Pr}$, and the Lewis number $\mathrm{Le}$. 
  The following two columns provide the far-field temperature and the minimum buoyancy for the reader's convenience. 
  The far-field temperature, $T_\infty$, is determined from Eq. (\ref{eq_salinity_interfacial_quadratic_a}) with $S_\infty=34~\textrm{g}~\textrm{kg}^{-1}$ and $R=2.3$ at $\textrm{Le}=200$. 
  The minimum buoyancy is $b_m = b(S_\infty,T_\textrm{i})$. 
  The columns 6--9 characterise the turbulent system in its stage of final simulation time.
  The simulations reach a boundary-layer height, $z_\textrm{est}$, of up to 0.19~m, and the turbulence intensities, $\frac{w_* z_\textrm{est}}{\nu}$, of up to 350, and $Re_\textrm{turb}=k^2/\left(\varepsilon \nu\right)$ of up to 25, 
    with turbulent kinetic energy $k$, viscous dissipation rate $\varepsilon$ and viscosity $\nu$. 
  The convective velocity scale, $w_*$, is defined as $w_*^3=\int_0^\infty \mathcal{H}\left(\langle b^\prime \upsilon_3^\prime \rangle\right) \langle b^\prime \upsilon_3^\prime \rangle \mathrm{d}z$, where $\mathcal{H}$ is the Heavyside function. 
  The last column is the resulting boundary thickness ratio of the simulation, $R$.
  All simulations have been initialised with an initial boundary thickness ratio of $R=2$. 
  The grid size of the simulations is $2560\times1152\times2560$. 
  At final time of the simulations the boundary layer reaches an aspect ratio between 4:1 and 5:1. 
  }
\label{tab_simulations_sbm}
\end{table*}
\end{center}

\section{The Flux Ratio}
\label{sec_sbm_interface}
From the full expressions of the heat flux and the salt flux,
\begin{linenomath}
\begin{subequations}
\label{eq_fluxes}
\begin{align}
&F_\textrm{h}(z,t) = \rho_\textrm{water} c_p \left(\kappa_\textrm{t}\partial_3 \left\langle T \right\rangle\left(z,t\right) - \left\langle \upsilon_3^\prime  T^\prime \right\rangle\left(z,t\right) \right) \textrm{ and} \\
&F_\textrm{s}(z,t) = \rho_\textrm{water} ~~~~    \left(\kappa_\textrm{s} \partial_3 \left\langle S \right\rangle\left(z,t\right) - \left\langle \upsilon_3^\prime  S^\prime \right\rangle\left(z,t\right) \right) \textrm{,}
\end{align}
\end{subequations}
\end{linenomath}
one obtains the flux ratio, 
\begin{linenomath}
\begin{align}
&\gamma(z,t) = \frac{\kappa_\textrm{t}\partial_3 \left\langle \theta \right\rangle(z,t) - \left\langle \upsilon_3^\prime \theta^\prime \right\rangle(z,t)}{\kappa_\textrm{s} \partial_3 \left\langle \sigma \right\rangle(z,t) - \left\langle \upsilon_3^\prime  \sigma^\prime \right\rangle(z,t)} \textrm{,}
\label{eq_fluxratio}
\end{align}
\end{linenomath}
with the normalised temperature, $\theta$, and the normalised salinity, $\sigma$. 

\cite{sirevaag_turbulent_2009} uses only the turbulent contributions of Eq. (\ref{eq_fluxratio}) to determine the flux ratio: 
\begin{linenomath}
\begin{align}
\label{eq_fluxratio_turb}
&\gamma_\textrm{turb} = \frac{\left\langle \upsilon_3^\prime  \theta^\prime \right\rangle\left(z,t\right)}{\left\langle \upsilon_3^\prime  \sigma^\prime \right\rangle\left(z,t\right)} \textrm{.}
\end{align}
\end{linenomath}
This turbulent flux ratio is identical to the ratio of turbulent exchange coefficients of \cite{notz_impact_2003} and \cite{mcphee_revisiting_2008} if the interface flux ratio is approximated by the mean turbulent flux ratio. 
\cite{gade_when_1993} uses only the molecular contributions of Eq. (\ref{eq_fluxratio}) at the wall to determine the interface conditions. 
This molecular-flux ratio is
\begin{linenomath}
\begin{align}
\label{eq_fluxratio_mol}
&\gamma_\textrm{mol}  = \textrm{Le} \frac{\partial_3 \left\langle \theta \right\rangle\left(z,t\right)}{\partial_3 \left\langle \sigma \right\rangle\left(z,t\right)}.
\end{align}
\end{linenomath}
The approximation to reduce the flux ratio either to $\gamma_\textrm{mol}$ at the wall or $\gamma_\textrm{turb}$ in the outer layer is reasonable, because there the corresponding contributions to the fluxes dominate numerator and denominator of $\gamma$ (see Figure \ref{fig_gamma_y}b). 

The determination of the interface conditions, however, requires that $\gamma$ is evaluated at the interface, where $\gamma_\textrm{i} \equiv \left. \gamma \right|_{z_\textrm{i}} = \textrm{Le}~R^{-1}$ with boundary-thickness ratio $R$ [Eq. (\ref{eq_bl_thickness_ratio}]. 
The use of $\gamma_\textrm{turb}$ as a surrogate for $\gamma_\textrm{mol}$ is convenient, because it allows for the employment of turbulent-flux measurements from the field. 
It has been shown, however, that (in a volume averaged sense [indicated by $\langle \cdotp \rangle_V$]) $\langle \gamma_\textrm{turb} \rangle_V \neq \langle \gamma_\textrm{mol} \rangle_V$ for the double-diffusive regimes of diffusion \citep{gargett_direct_2003}, saltfingering \citep{kimura_turbulence_2011}, and semi-convection \citep{kupka_semi-convection_2015}. 
We infer this from the temporal evolution of effective diffusivities of temperature, $K_\textrm{t}$, and salinity, $K_\textrm{s}$, given therein. 
According to Eqs.~(\ref{eq_fluxratio_turb},~\ref{eq_fluxratio_mol}), 
\begin{linenomath}
\label{eq_le_turb}
\begin{align}
&\left\langle \frac{K_\textrm{t}}{K_\textrm{s}} \right\rangle_V= \left\langle \frac{\langle \upsilon^\prime T^\prime \rangle}{\partial_3 \langle T \rangle} \frac{\partial_3 \langle S \rangle}{\langle \upsilon^\prime S^\prime \rangle} \right\rangle_V = \left\langle \frac{\gamma_\textrm{turb}}{\gamma_\textrm{mol}} \right\rangle_V~\textrm{Le} \textrm{,}
\end{align}
\end{linenomath}
but in the given references, $\left\langle K_\textrm{t}/K_\textrm{s} \right\rangle_V \neq \textrm{Le}$, and therefore $\langle \gamma_\textrm{turb} / \gamma_\textrm{mol} \rangle_V \neq 1$. 
With our simulations we support this finding for an extended range of $R_\rho^s$ for the semiconvective regime and we investigate the diffusive and the turbulent contribution to the flux ratio separately. 

Even though $\gamma_\textrm{turb} \neq \gamma_\textrm{mol}$, it might still be valid to employ turbulent-flux measurements once the relation between turbulent and molecular flux ratio is clarified. 
In the following, we explore the vertical structure of $\gamma$ from our simulations to understand the relation between turbulent and molecular flux ratio. 

\subsection{Vertical Structure of the Flux Ratio}
\begin{figure*}[tbp]
\begin{center}
\includegraphics[width=.6\linewidth]{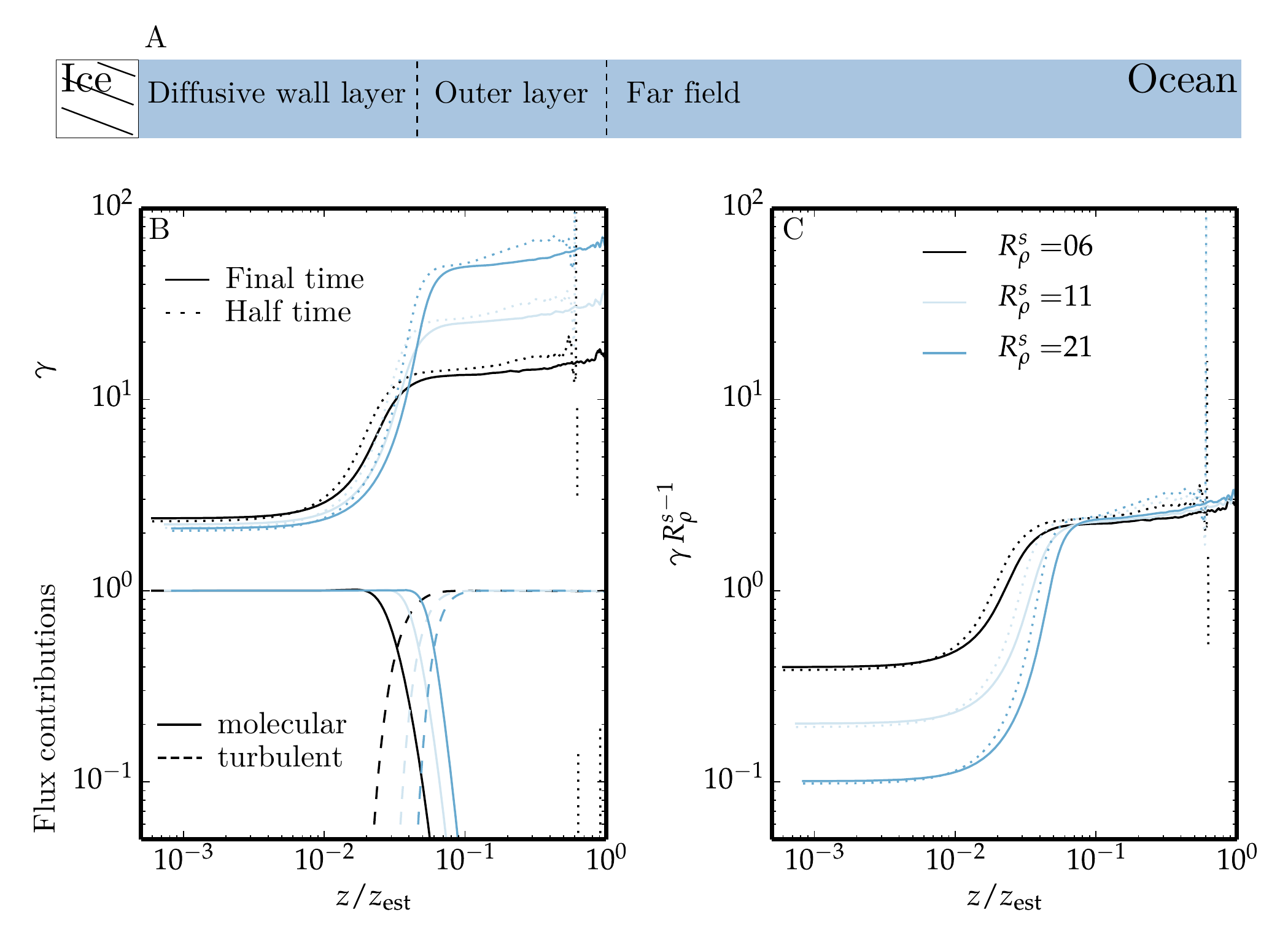}
\caption{\small{
a) Illustration of the vertical structure of free convection beneath a melting ice--ocean interface in terms of the abscissa of figure b. 
b) Molecular fraction of total heat transport (solid line on the scale~$< 1.0$) and turbulent fraction of total heat transport (dashed line on the scale~$< 1.0$). 
   Ratio between the total heat transport (molecular + turbulent) and the total transport of salinity, $\gamma$, at final simulation time (solid line on the scale~$> 1.0$) and at half simulation time (dotted line on the scale~$> 1.0$). 
   This figure shows that $\gamma_\textrm{turb}$ describes this spatial structure well in the outer layer, $\gamma_\textrm{mol}$ describes it well next to the interface. 
c) Ratio $\gamma$ at final simulation time, but scaled with $R_\rho^s$. 
   Colours indicate different density ratios, $R_\rho^s$. 
   The flux ratio fluctuates wildly (vertical lines) between the outer layer and the far field where there is hardly any turbulent nor molecular fluxes anymore. 
}}
\label{fig_gamma_y} 
\end{center}
\end{figure*}

The vertical structure of free convection beneath a melting, horizontal, smooth ice--ocean interface in the semiconvective regime is well described by a two-layered structure: 
  a diffusion-dominated wall layer (the diffusive wall layer) and a turbulence-dominated outer layer (in sea-ice literature sometimes referred to as mixed layer, $_\textrm{ml}$) (see Figure \ref{fig_gamma_y}a). 
Molecular diffusion dominates the heat transport next to the interface (see Figure \ref{fig_gamma_y}, solid line on the scale~$< 1.0$). 
Turbulent transport dominates the heat transport in the outer layer (see Figure \ref{fig_gamma_y}, dashed line on the scale~$< 1.0$). 
This description is similar to the two-layered structure of the ice--lake system 
 and is inspired by the more detailed description of the boundary layer of free convection over a heated plate in \cite{mellado_direct_2012}. 

In the diffusive wall layer, we observe a significantly smaller flux ratio than in the outer layer (see Figure \ref{fig_gamma_y}b, solid lines on the scale~$> 1.0$). 
From the definition of the molecular flux ratio, $\gamma_\textrm{mol}$ [see Eq. (\ref{eq_fluxratio_mol})], one expects it to scale with $\textrm{Le}$ and $R^{-1}$. 
For our simulations at fixed Lewis number, $\textrm{Le}$, the flux ratio seems steady (cf. profiles at half time of the simulation, dotted lines) and independent of the far-field conditions (represented by varying $R_\rho^s$). 
We find a boundary thickness ratio $R=\textrm{Le}~\gamma_\textrm{i}^{-1}\approx1.7$--$1.9$ (see Table \ref{tab_simulations_sbm}). 
This constant $R$ respresents a sequential layering of mean-temperature and mean-salinity profiles at the interface.  
The visualisations show, however, how temperature-driven and salinity-driven buoyancy fluctuations alternate (see Figure \ref{fig_vis}). 

In the outer layer, we observe a larger flux ratio than in the diffusive wall layer (see Figure \ref{fig_gamma_y}b, solid lines on the scale~$> 1.0$). 
The observed flux ratio depends on the far-field conditions (cf. varying flux ratio for varying $R_\rho^s$). 
From the definition of the turbulent flux ratio, $\gamma_\textrm{turb}$ [see Eq. (\ref{eq_fluxratio_turb})], one expects it to be independent of the diffusivity ratio, $\textrm{Le}$, and to scale with a certain temperature--salinity ratio.  
If the outer layer is well-mixed, temperature and salinity will be distributed homogeneously. 
Their mixing is driven by a buoyancy-reversal instability. 
According to Eq. (\ref{eq_b_bilinear_nd}), the buoyancy-reversal instability is favoured by a temperature--salinity ratio commensurate with the density ratio, $R_\rho^s$, i.e. $\theta / \sigma \sim R_\rho^s$ (see supplementary material). 
Consequently, temperature and salinity must be entrained in a proportion that is commensurate with this $R_\rho^s$ to maintain this $\theta / \sigma$. 
Our simulations support this argument and provide the scaling
\begin{linenomath}
\begin{align}
\gamma_\textrm{turb} = 2.3 ~ R_\rho^s  
\label{eq_gamma_turb_const}
\end{align}
\end{linenomath}
(see Figure \ref{fig_gamma_y}c). 
Interestingly, the scaled turbulent flux ratio seems to yield a constant similar to that of the interface flux ratio. 

In summary, the vertical structure of the flux ratio is well described by two parameters. 
A boundary thickness ratio, $R$, describes the molecular flux ratio at the interface. 
A density ratio, $R_\rho^s$, describes the turbulent flux ratio in the outer layer. 
Because $R_\rho^s$ is an independent control parameter that defines the setup, only $R$ remains to be determined to know the vertical structure of the flux ratio.

\subsection{Temporal Evolution of the Fluxes and their Ratio}
The vertical structure of the flux ratio results from the vertical structure of the temperature and salinity flux. 
All of our simulations exhibit a similar vertical flux structure: the molecular fluxes at the interface are higher than the turbulent fluxes in the outer layer (not shown). 
As a consequence of this vertical flux structure, the flux profiles are not steady. 
The water in the diffusive wall layer continuously cools and freshens and thereby decreases the molecular fluxes of temperature and salinity at the interface. 

While the molecular fluxes at the interface keep decreasing in time, we observe a quasi-steady flux ratio that is described by $R$ and $\gamma_\textrm{turb}$. 
Accordingly, we observe that the turbulent fluxes decrease along with the molecular fluxes for all of our simulations. 
The total buoyancy flux, the integral of molecular and turbulent fluxes, decreases proportional to the growth of the boundary layer (not shown). 
Consequently, the turbulence will finally cease. 

For externally forced turbulence, the behaviour of the flux ratio remains an open question. 
However, all indication from our simulations is that the mixing in the outer layer will not affect the boundary thickness ratio, $R$.

\section{Discussion}

\begin{figure*}[tbp]
\begin{center}
\includegraphics[width=.8\linewidth]{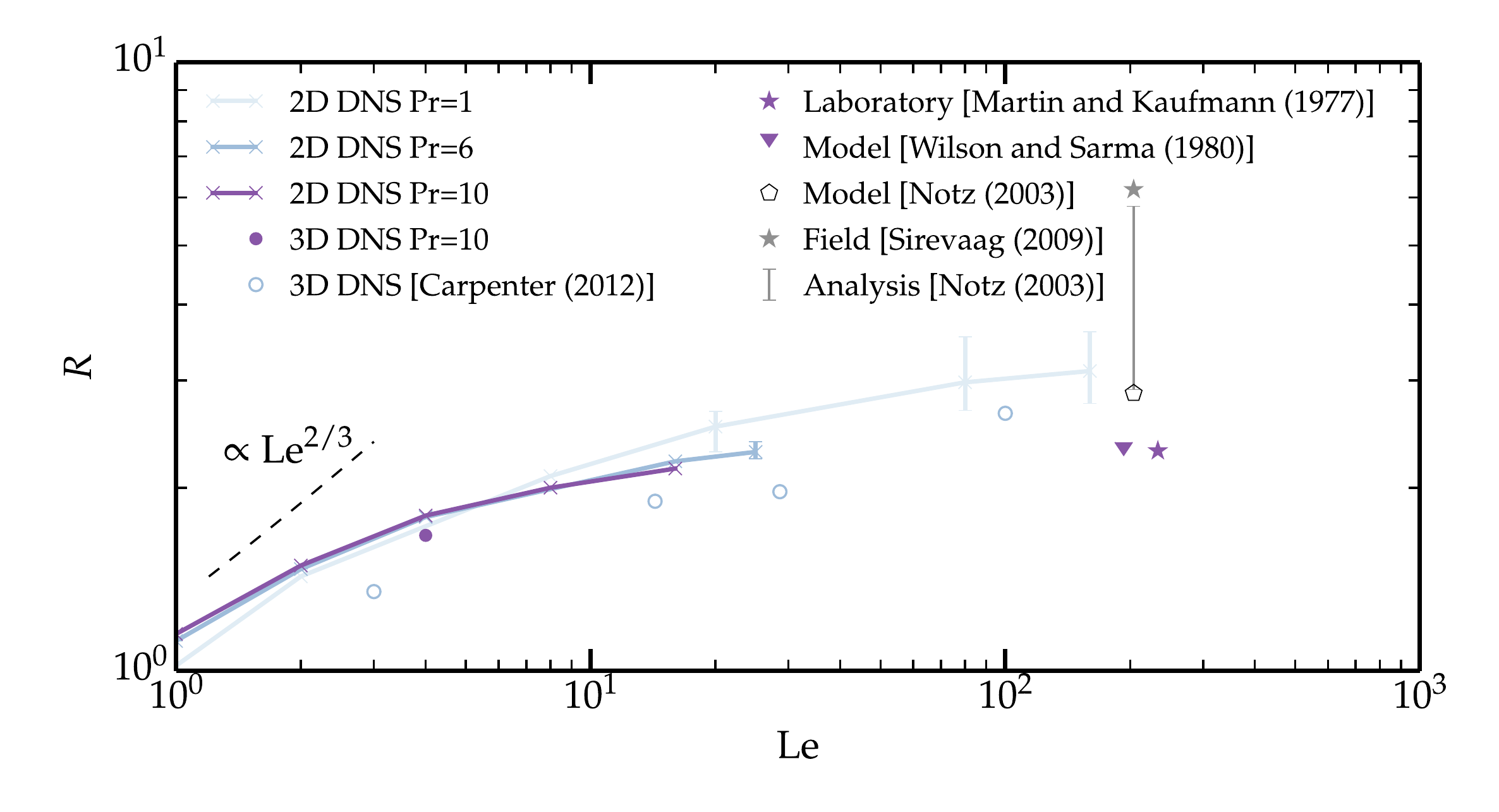}
\caption{\small{
Lewis-number series of boundary thickness ratio, $R=\delta_\textrm{t}/\delta_\textrm{s}$. 
Colors indicate different Prandtl numbers: $\textrm{Pr}=1$ (light blue), $\textrm{Pr}=6.25$ (blue), $\textrm{Pr}=10$ (purple). 
Our direct numerical simulations are of $R_\rho^s=6$ and $R_\textrm{init}\approx2$. 
The error estimations for the direct numerical simulations is the maximum and minimum of the temporal fluctuations around the mean boundary thickness ratio. 
The dashed line indicates the expected behaviour of R from the diffusive scalings of free convection over a heated plate. 
The data of $\textrm{Pr}=10$ (purple) suggests a boundary thickness ratio of 2--2.3.
}}
\label{fig_R} 
\end{center}
\end{figure*}

In this work, we have so far determined $R$ from simulations at fixed $\textrm{Pr}=10$ and fixed $\textrm{Le}=4$. 
$R$ can be described by a constant---almost independent of the far-field conditions and also independent of varying initial conditions (see supplementary material). 
We will now use former studies along with a series of two-dimensional simulations for varying Prandtl number and Lewis number to assess the boundary thickness ratio at as high a Lewis number as it occurs in the Arctic Ocean. 

The mean evolution of two-dimensional turbulence does not represent the mean evolution of three-dimensional systems correctly. 
Nonlinear flow phenomena, such as cascades, coherent structures and dissipative processes, however, take place in both systems and a common conceptual framework between two- and three-dimensional turbulence exists \citep{tabeling_two-dimensional_2002}. 
It can hence be instructive to employ simulations of two-dimensional turbulence \citep{fedorovich_atmospheric_2004} in double-diffusive systems \citep{zweigle_direkte_2011}. 
Moreover, they appear to yield an energy balance at the interface that is to within 10\% accuracy to that of three-dimensional simulations. 
This has already been observed for simulations of the ice--lake system (not shown), 
  and it is so for the simulations of the ice--ocean system in both temperature and salinity. 
\cite{carpenter_simulations_2012} have shown that two-dimensional direct numerical simulation \textquotedblleft accurately \textquotedblright captures the heat flux and interfacial structures of three-dimensional direct numerical simulations when the density variation due to salinity is at least three times larger than the density variation due to temperature. 

From the collection of two-dimensional and three-dimensional simulations (see Figure \ref{fig_R}), it emerges that $R$ increases with increasing Lewis number, $\textrm{Le}$. 
In other words, the larger the difference of temperature diffusivity and salinity diffusivity is, the larger the difference in their gradient thicknesses. 
The diffusivity of the scalar that defines the buoyancy, $\kappa$, is a key parameter in the diffusive wall layer in free-convective flows next to a Dirichlet interface. 
The corresponding gradient thickness, $\delta$, scales as $\kappa^{2/3}$ \citep{mellado_direct_2012}. 
One might therefore expect the ratio of two gradient thicknesses which are controlled by different diffusivities to scale with the ratio of diffusivities as ${\left( \kappa_\textrm{t} / \kappa_\textrm{s}\right)}^{2/3}$ as long as the two scalars do not couple to each other (see Figure \ref{fig_R}, dashed line). 

Interestingly, $R$ levels off with further increasing $\textrm{Le}$ indicating that the temperature and salinity field interact with each other (see Figure \ref{fig_R}). 
\cite{carpenter_simulations_2012} reason that the temperature profile effectively feels the salinity interface as a solid conducting plane once $\textrm{Le}$ is high enough. 
Then, the development of a temperature sublayer at the felt salinity interface should only follow the Prandtl number.  
For a fixed Prandtl number, it is hence reasonable to expect that the boundary thickness ratio approaches a constant. 

\begin{figure}[tbp]
\begin{center}
\includegraphics[width=.8\linewidth]{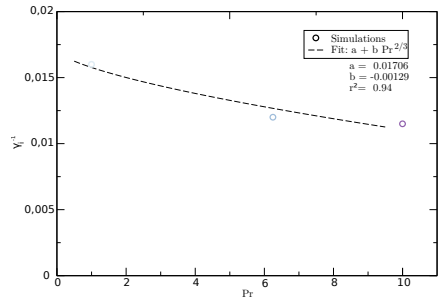}
\caption{\small{
Simulated $\gamma$ as extrapolated from Figure \ref{fig_R} for $\textrm{Le}\approx200$ (circles) and power-law prediction by \cite{mcphee_dynamics_1987} (dashed line).
}}
\label{fig_gamma_Pr} 
\end{center}
\end{figure}
In summary, $R$ becomes independent of $\textrm{Le}$ once $\textrm{Le}$ crosses a critical value. 
We observe that $R$ still depends on the diffusivity of the more strongly diffusing scalar, temperature. 
Our simulations are consistent with the scaling, $R \propto \textrm{Pr}^{2/3}$, suggested by \cite{mcphee_dynamics_1987} (see Figure \ref{fig_gamma_Pr}). 
For $\textrm{Pr}=10$, a $\textrm{Le}$ of four already seems to be high enough to approach independence of $R$ on $\textrm{Le}$. 
From our simulation we measure $R=$1.7--1.9. 
That yields $\gamma_\textrm{mol}=$105--118. 

\subsection{Comparison to Former Studies}
Several previous studies had independently targeted the boundary thickness ratio, $R$, beneath an ice--ocean interface. 

\cite{notz_impact_2003} assessed the bulk flux ratio by modeling observations. 
They use this bulk flux ratio to determine the interface temperature and salinity. 
In this sense they equate $\gamma_\textrm{bulk}$ with $\gamma_\textrm{i}$. 
This hypothetical equality implies a hypothetical boundary thickness ratio, $R$. 
The range they estimate for $\gamma_\textrm{bulk}$ corresponds to a $R$ range of [$2.9 < R < 5.7$] (see Figure \ref{fig_R}, grey bar). 
However, their hint that lower $R$ values fit the data better already points towards the smaller value obtained in this study. 

\cite{carpenter_simulations_2012} assessed the flux ratio of a fluid--fluid interface---warm and fresh fluid on top of cold and salty fluid. 
Buoyancy-reversal instabilities on both sides of a sharp fluid--fluid interface promote free convection. 
They find that turbulence is not able to penetrate the stable stratification of the interface core. 
Just like a rigid interface, their fluid--fluid interface is dominated by molecular fluxes. 
They observe a boundary thickness ratio, $R$, of 2.5 at $\textrm{Pr}=6.25,~\textrm{Le}=100,~R_\rho^s=6$, (see Figure \ref{fig_R}, circle) 
similar to the values observed in our simulations. 

\cite{gade_when_1993} assessed the molecular flux ratio from the boundary thickness ratio of interfacial profiles of temperature and salinity. 
He finds $R=2.3$ from the laboratory work of \cite{martin_experimental_1977} (see Figure \ref{fig_R}, purple star) and $R=2.26$ from the modeling work of \cite{wilson_prediction_1980} (see Figure \ref{fig_R}, pyramid). 

\cite{sirevaag_turbulent_2009} assessed the turbulent flux ratio by field measurements. 
Depending on the time span of the temporal averaging and the threshold criteria that are applied to the measurement data, he determines turbulent flux ratios between 23 and 37 in case of rapid melting. 
These turbulent fluxes are well within the range of those that we observe in the outer layer of our simulations (see Figure \ref{fig_gamma_y}b). 
The turbulent flux in the outer layer, however, is not a good proxy the flux ratio at the interface. 

The outer-layer flux ratio determined in this work does agree with field measurements. 
Regarding the interface flux ratio, we estimate a boundary thickness ratio of
\begin{linenomath}
\label{eq_bl_thickness_ratio_determination}
\begin{align}
R=\left(2.2\pm0.2\right)
\end{align}
\end{linenomath}
from all of the evidence presented so far: 
from the trends of all two-dimensional simulations to lose their dependence on the Lewis number (see Figure \ref{fig_R}, lines), 
from the value which the two-dimensional simulations approach for increasing Lewis number at $\textrm{Pr}=10$ (purple line), 
from the laboratory-experiment value of \cite{martin_experimental_1977} (purple star), 
and from the fact that the three-dimensional simulation yields a slightly smaller value than does the two-dimensional simulation (purple dot). 
This interval yields an interface flux ratio in the range of 83 to 100. 
The interface flux ratio determined in this work does agree with that obtained from laboratory work, modeling work and numerical work of similar setups. 

The interface flux ratio has been suspected to depend on the surface roughness of the ice--ocean interface \citep{notz_impact_2003}. 
Our investigations of a smooth ice--ocean interface show that no surface roughness at all is needed to reproduce the turbulent flux ratios that have been suggested as reasonable. 
 
\subsection{Dependence of the Boundary Thickness Ratio on Far-field Temperature and Salinity}
\begin{figure}[tbp]
\begin{center}
  \includegraphics[width=.9\linewidth]{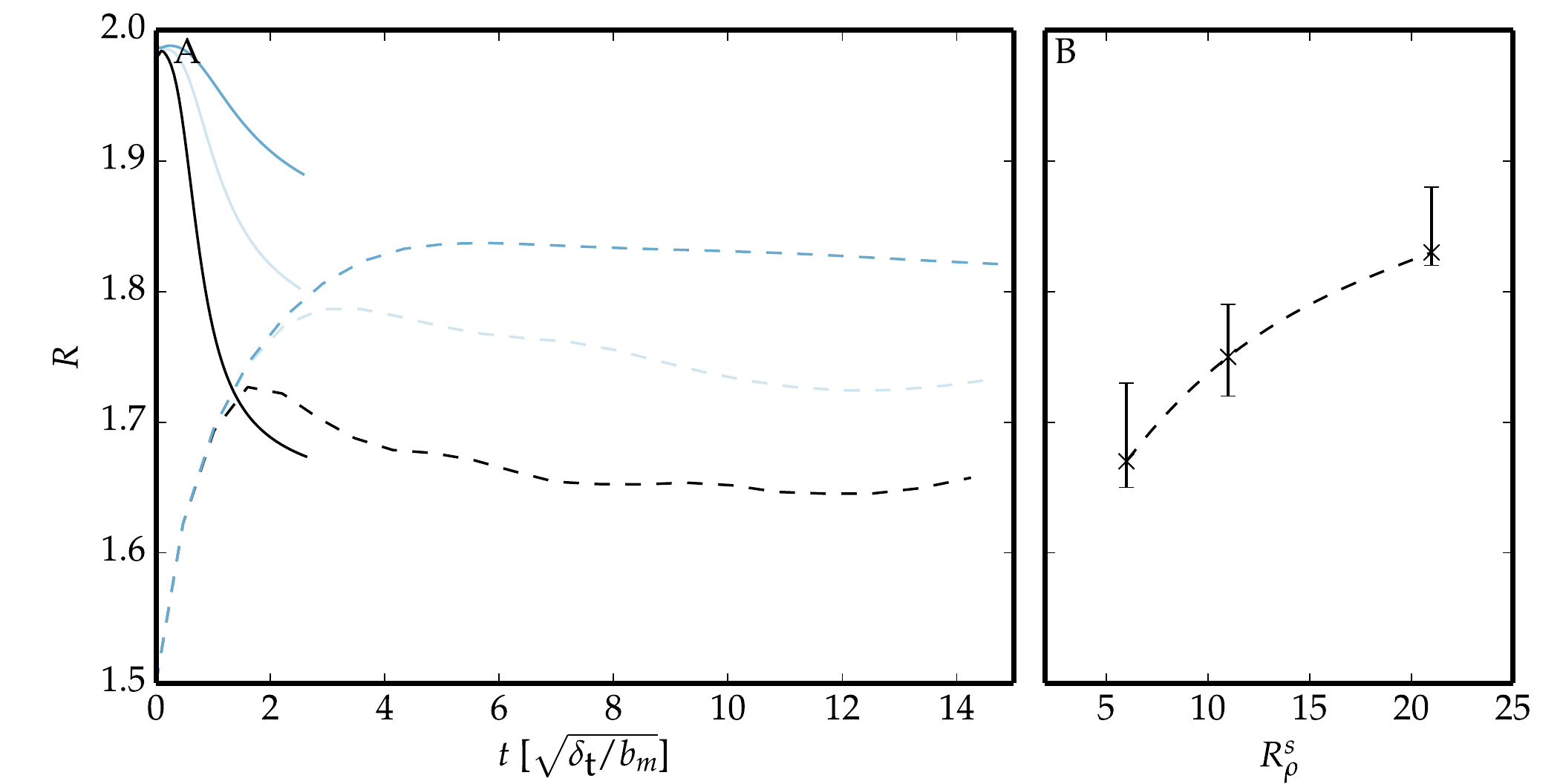}
\caption[]{\small{
a) Evolution of the boundary thickness ratio in time. Three-dimensional simulations (solid lines) approach the values obtained from two-dimensional simulations (dashed lines). 
Colours indicate the density ratio of the simulation as given in Figure \ref{fig_gamma_y}c. 
b) Density-ratio series of the boundary thickness ratio.
}} 
\label{fig_R_R}
\end{center}
\end{figure} 

So far it seems that the boundary thickness ratio is a universal property of a melting ice--ocean interface. 
However, a weak dependence on the far-field conditions remains. 
As explained in section two, the far-field conditions reflect in the overall temperature range, $T_\textrm{i}-T_\infty$, and in the overall salinity range, $S_\textrm{i}-S_\infty$, of the system. 
The independent control parameter that characterises the far-field conditions is the density ratio, $R_\rho^s$. 
The dependence of $R$ on $R_\rho^s$ is illustrated in Figure \ref{fig_R_R}. 
The variation in $R$ accounts for about 10\% of its value. 
The dependence is faint and seems to level off for increasing $R_\rho^s$.

\subsection{The Melt Rate}
\begin{figure*}[tbp]
\begin{center}
  \includegraphics[width=.9\linewidth]{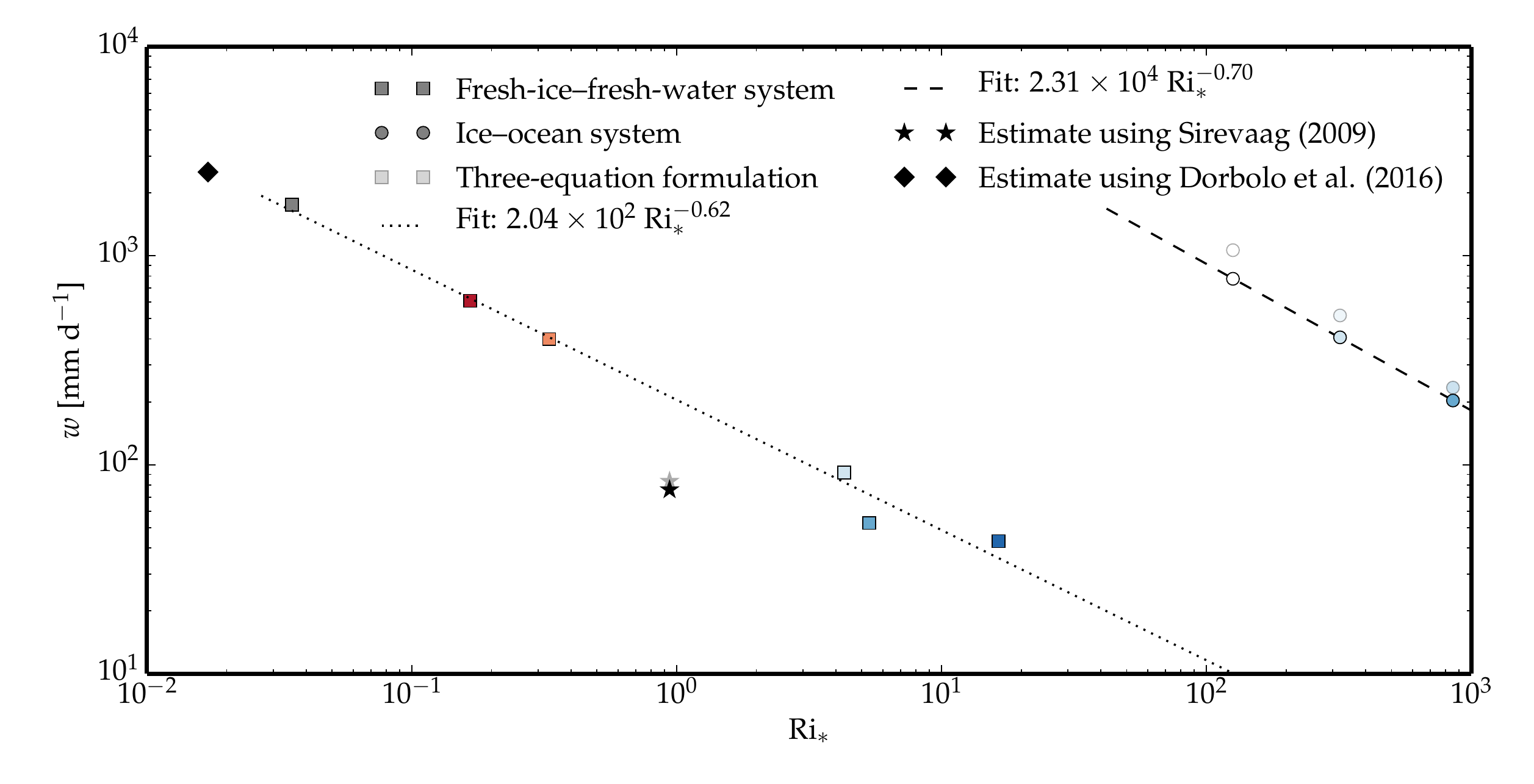}
\caption[]{\small{
Melt rates, $w$, over convective Richardson number, $\textrm{Ri}_*$. 
The convective Richardson numbers of the simulations are determined from the strength and apparent position of the buoyancy-reversal instability and the convective velocity scale (filled circles). 
The convective Richardson number of the ice--lake system are taken from \cite{keitzl_impact_2016} (filled squares). 
Colours indicate the density ratio, $R_\rho^s$, and Richardson number, $\textrm{Ri}_0$, of the setups as given in Figure \ref{fig_gamma_y}c and in Fig. 1/Tab. 1 of \cite{keitzl_impact_2016}. 
The melt rates of the ice--lake system agree well with independent laboratory measurements of \cite{dorbolo_rotation_2016} (black diamond). 
They measured the mass flux of a rotating ice disc melting in fresh water. 
From PIV measurements, they found convective velocities of $10$~mm~s$^{-1}$ at a far-field temperature of 20~\textcelsius~(personal communication). 
The turbulent-instrument-cluster measurements of \cite{sirevaag_turbulent_2009} follow more closely our simulations of the ice--lake system than those of the ice--ocean sytem (black star). 
His measurements of far-field conditions and friction velocity allow us to reconstruct a convective Richardson numbers, where we ued $R=2.15$. 
We assess his measurement in the discussion section. 
The three--equation formulation with $\gamma_\textrm{i}=90$ and with the commonly accepted value of $\alpha_\textrm{h}\approx0.01$ overestimates the melt rates when the convective velocity scale is used to mimic the friction velocity of forced convection (shaded symbols). 
}}
\label{fig_meltrate}
\end{center}
\end{figure*} 

The determination of the melt rate from the three--equation formulation requires knowledge of two parameters:
the bulk heat exchange coefficient, $\alpha_\textrm{h}$, and the boundary thickness ratio, $R$. 
The latter is relevant for the determination of the interface temperature, $T_\textrm{i}$. 
The melt rate is then given as a function of far-field temperature, $T_\infty$, and friction velocity, $u_\textrm{*0}$ (see Eq. (\ref{eq_alpha_h}).
In this work, we have determined $R$ to be $2.2\pm0.2$. 
In the absence of temperature and salinity fluxes in the ice, the three--equation formulation yields melt rates that are up to 40~\% lower than previous estimates due to the influence of $R$ on the interface temperature (see Figure \ref{fig_dev}b). 

With the interface temperature available, it proves convenient to provide melt rates as a function of the convective Richardson number,
\begin{linenomath}
\label{eq_richardson_convective}
\begin{align}
  \textrm{Ri}_* = \frac{\Delta b z_0}{w_*^2} \textrm{,}
\end{align}
\end{linenomath}
  with the strength of the stable stratification, $\Delta b = b\left(z_\textrm{i}\right) - b_m$,
  the thickness of the diffusive wall layer, $z_0$, 
  and the convective velocity scale, $w_*$ (see Figure \ref{fig_meltrate}). 
The convective Richardson number describes the ratio between the kinetic energy that a fluid particle needs to overcome the diffusive shield beneath the ice and its kinetic energy. 
A comparison of the Richardson-number dependence of the melt rates measured from our simulation (see Figure \ref{fig_meltrate}, filled circles) to that of the ice--lake system (see Figure \ref{fig_meltrate}, filled squares), shows how both systems follow a similar working principle. 
As the far-field temperature increases, the buoyancy-reversal instability strengthens and the convective Richardson number decreases (see Figure \ref{fig_meltrate}, filled circles). 
With increasing convective motion beneath the ice, the melt rates of the ice increase. 
Compared to the ice--ocean system, the ice--lake system has larger an extent of the diffusive wall layer but significantly weaker a diffusive shield at similar free-convection velocities for systems of similar size. 
Consequently, the Richardson numbers of the ice--lake system are smaller by a factor of 100. 
In this case, convective motions are strong enough to overcome the diffusive shield. 
In the case of ice--ocean free convection, however, we have observed that convective motions were not strong enough to overcome the diffusive shield because
turbulence is statistically unsteady and will finally cease. 

The melt rates in conditions of forced- and mixed-convection remain to be ascertained. 
\cite{sirevaag_turbulent_2009} provides a well-controlled field measurement of mixed convection in ice--melting conditions. 
He measured the far-field temperature ($-0.86$~\textcelsius), the far-field salinity ($34.4~\textrm{psu}$), the heat flux ($268~\textrm{W}~\textrm{m}^{-2}$) and the friction velocity ($0.9 \times 10^{-2}~\textrm{m}~\textrm{s}^{-1}$). 
From his measurements, we reproduce the interface temperature ($-1.27$~\textcelsius), the interface salinity ($23.5~\textrm{psu}$), the buoyancy shielding $\Delta b$ ($8.8~\times10^{-2}~\textrm{m}~\textrm{s}^{-2}$), and the strength of the buoyancy-reversal instability $b_m$ ($1.5\times10^{-4}~\textrm{m}~\textrm{s}^{-2}$).
If his heat-flux measurement under the given mixed-convection conditions is representative for the diffusive flux at the interface, we estimate $z_0=0.9 \times 10^{-3}~\textrm{m}$. 
In the absence of fluxes within the ice, his measurements yield a melt rate of $76~\textrm{mm}~\textrm{d}^{-1}$ at a convective Richardson number of about 1. 
His measurement deviates thus from the Richardson-number dependence seen for the simulated ice--ocean system and follows more closely the simulated ice--lake system. 
From this measurements, we learn that mixed convection occurs at lower Richardson numbers because convective motions eventually compete with the diffusive shield just as in ice--lake free convection. 
From the strength of the buoyancy-reversal instability, $b_m$, and the stably stratified buoyancy shield, $\Delta b$, one expects a reference Richardson number, $\textrm{Ri}_0=\Delta b / \left| b_m \right| + 1$, for his measurement of about 586 [see Eq. (\ref{eq_richardson})].
If the measurement were made under free-convection conditions, one further expects a turbulent flux ratio of about $1350$ [from Eq. (\ref{eq_gamma_turb_const})]. 
Instead \cite{sirevaag_turbulent_2009} found $\gamma_\textrm{turb}=33$. 
We conjecture that the influence of external forcing tends to even out differences between the turbulent fluxes of heat and salt. 
The role of the mixing fraction, $R_\rho^s$, in sustaining the convective motion decreases with the strength of the shear. 
In the extreme of forced convection, one might hence expect a turbulent flux ratio that is not influenced by $R_\rho^s$ anymore and approaches unity.

\subsection{The turbulent flux ratio under ice--growing conditions}

Even though we have considered a melting--ice scenario for our studies, 
  we now assess the turbulent flux ratio of a growing--ice scenario. 
\cite{mcphee_revisiting_2008} study the double-diffusive tendencies under growing ice. 
They found an asymmetric behaviour of the double-diffusive process between freezing and melting ice: 
While turbulent flux ratios measured in melting--ice scenarios are for example of order $10^1$--$10^2$, the turbulent flux ratio in growing--ice scenarios was assessed to be unity. 
As opposed to melting ice, they suggest that for growing ice the double-diffusive tendencies are relieved by dynamics within the mushy layer above the advancing ice front. 

This mushy layer prevails for growing ice because the salt of the ocean water is embedded between growing ice crystals. 
As long as ice continues to grow, the lower most advancing front of bulk ice will always embed salt corresponding to the salinity of the ocean water as it grows \citep{notz_desalination_2009}.
Therefore, we do not expect any salt gradient between the ocean water and the lower most frontDesalination of the mushy layer. 
The interface salinity is that of the ocean water. 

The presence of an interface salinity lower than the salinity in the outer layer is a prerequisite for the promotion of the buoyancy-reversal instability. 
In the absence of the buoyancy-reversal instability (the forcing mechanism of free convection), the density ratio, $R_\rho^s$, does no longer control nor influence the ratio in which salinity and temperature mix in the outer layer. 
The forcing mechanisms that remain are for example vertical shear from draining salt plumes and horizontal shear due to drifting of ice. 
A shear forcing, however, does not distinguish between the mixing of temperature and salinity, because---by definition---the diffusivities do not influence the turbulent flux ratio [see Eq. (\ref{eq_fluxratio_turb})]. 
Based on our findings, we endorse the argument of \cite{mcphee_revisiting_2008}; 
Not dynamics within the mushy layer but its mere presence implies a turbulent flux ratio of unity.

\section{Conclusions}
The main objective of this study has been the determination of the ratio of heat and salt flux. 
This flux ratio is used in models of ice--ocean interaction to control the interface conditions and thus also the melt rates. 
We have obtained the vertical structure of the flux ratio from simulations of free convection beneath a melting, horizontal, smooth ice--ocean interface in the semiconvective regime. 
By means of direct numerical simulations, we have determined the flux ratio for the first time not only in the outer layer but also directly at the interface. 
We have reported two main findings.

First, the ratio of heat and salt flux has the following vertical structure: 
In the outer layer, the ratio depends strongly on the far-field temperature and salinity of the water. 
A commonly used independent control parameter, the density ratio, $R_\rho^s$, can be used to scale the flux ratio there.
The flux ratios obtained in field measurements from \cite{sirevaag_turbulent_2009} are easily reproduced for varying $R_\rho^s$. 
Next to the interface, however, the flux ratio becomes almost independent of the far-field conditions as has been indicated before by \cite{holland_modeling_1999}. 
The flux ratio has to be evaluated at the interface to obtain the value relevant for the determination of the interface conditions of the ice--ocean interface. 
Our simulations indicate that direct measurements of the interface flux ratio based on the turbulent fluxes will be difficult, because next to the interface the turbulent contribution ceases and is not a good proxy for the molecular contribution. 
Instead we have presented a consistent evaluation of the flux ratio based on the total heat and salt fluxes. 

Second, the interface flux ratio is three times as large as previously assessed based on turbulent-flux measurements at a certain distance from the interface. 
Instead of the currently accepted value of the flux ratio, $\gamma_\textrm{turb}=33$, which corresponds to the turbulent flux ratio in the outer layer, we find $\gamma_\textrm{i}\approx83$--$100$ to be more realistic at the interface. 
With $\gamma_\textrm{i}$ the interface conditions are determined according to Eq. (\ref{eq_bc1}),~Eq. (\ref{eq_salinity_interfacial_quadratic_a}),~and $\Delta T_\gamma=0.85~K$. 
Compared to our improved estimate, melt rates of the ice--ocean interface based on the three-equation formulation using the too low value $\gamma_\textrm{turb}=33$ are overestimated by up to 40\%.

\section*{Supplementary Material}

\subsection*{Dependence on Initial Conditions}
\label{sec_ic}

\begin{figure*}[tbp]
\begin{center}
  \includegraphics[width=.7\linewidth]{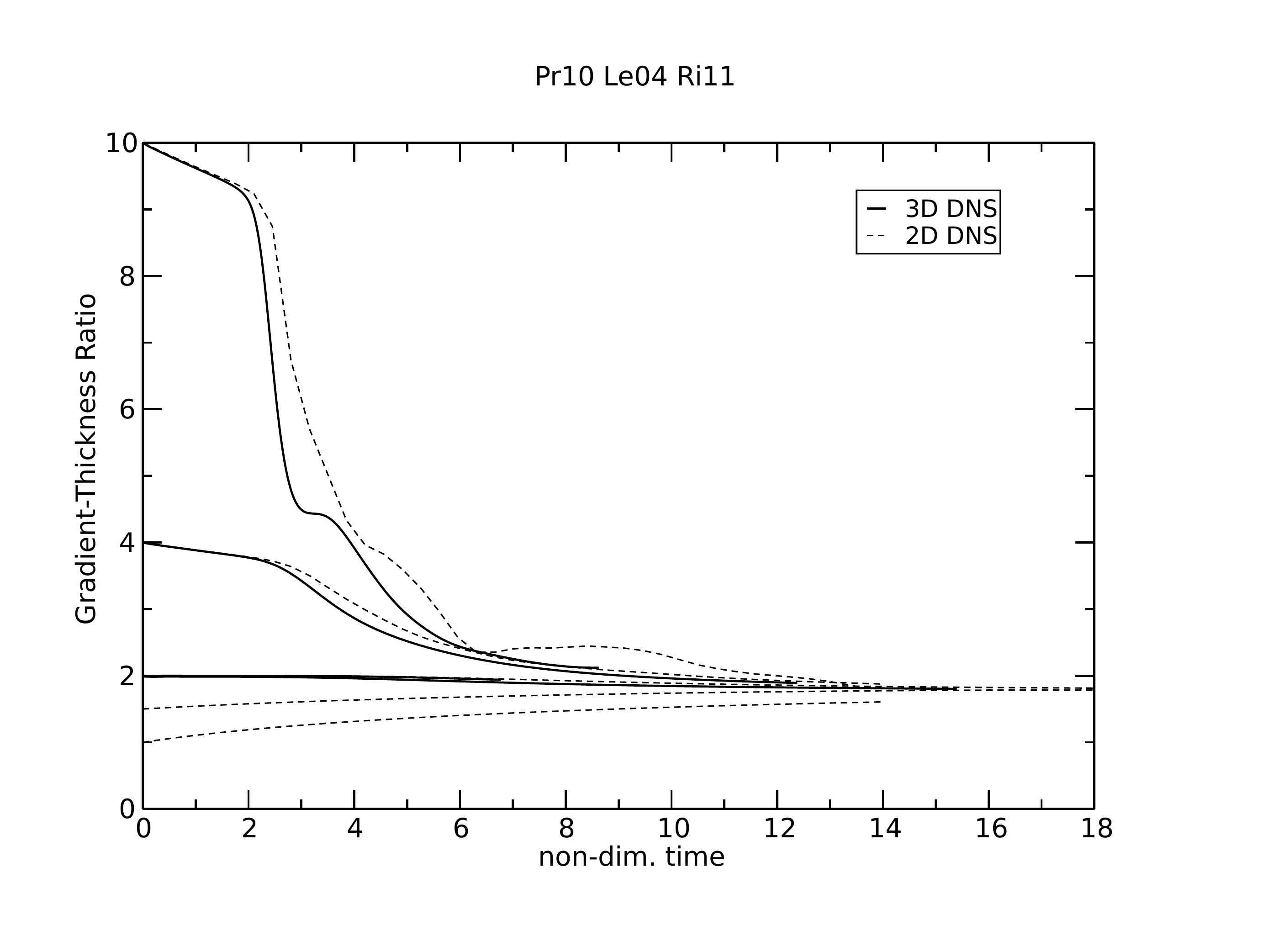}
\caption[]{\small{
Temporal evolution of simulations of varying initial conditions as given in Table \ref{tab_simulations_sbm_3d_ic} and as obtained from two-dimensional simulations. 
}}
\label{fig_R_time}
\end{center}
\end{figure*} 

\begin{center}
\begin{table*}[htbp]     
\begin{tabular*}{\linewidth}{@{\extracolsep{\fill}}cccccccccccc}
\hline
\hline
\rule[-6pt]{0pt}{23pt}	&	Case	&$R(t=0)$&$\sqrt{\textrm{TKE}_\textrm{max}}(t=0)$	&&&	$z_\textrm{est}$ [m]	&$\frac{w_* z_\textrm{est}}{\nu}$	&	$\frac{k^2}{\left(\varepsilon \nu\right)}$	&$R_\textrm{final}$	& \\
\hline
\rule[-6pt]{0pt}{21pt}	&	ic1	&	10	& 0				&&&	0.29	&	1538 				&	204			&	 2.12			&	\\
\rule[-6pt]{0pt}{21pt}	&	ic2	&	4	& 0				&&&	0.22	&	522				&	~42			&	 1.89			&	\\
\rule[-6pt]{0pt}{21pt}	&	ic3	&	2	& 0.03				&&&	0.17	&	245				&	18			&	 1.80			&	\\
\hline
\hline
\end{tabular*}  
\caption{Properties of the numerical simulations of the ice--ocean system for different initial conditions (ic). 
  All simulations are conducted at $\textrm{Pr}=10$, $\textrm{Le}=4$, and at $R_\rho^s=11$. 
  The columns 4--7 characterise the turbulent system in its stage of final simulation time.
  The simulations reach a boundary-layer height, $z_\textrm{est}$, of up to 0.29~m, and the turbulence intensities, $\frac{w_* z_\textrm{est}}{\nu}$, of up to 1500, and $Re_\textrm{turb}=k^2/\left(\varepsilon \nu\right)$ of up to 200, 
    with turbulent kinetic energy $k$, viscous dissipation rate $\varepsilon$ and viscosity $\nu$. 
  The convective velocity scale, $w_*$, is defined as $w_*^3=\int_0^\infty \mathcal{H}\left(\langle b^\prime \upsilon_3^\prime \rangle\right) \langle b^\prime \upsilon_3^\prime \rangle \mathrm{d}z$. 
  The last column is the resulting boundary thickness ratio of the simulation, $R$.
  The grid size of the ic1-simulation is $1024\times1152\times1024$, and $2560\times1152\times2560$ for the rest. 
  }
\label{tab_simulations_sbm_3d_ic}
\end{table*}
\end{center}

Turbulent structures are known to quickly lose the memory of their history and so of their initial conditions. 
In the following, we monitor if this is true for our property of interest, the boundary thickness ratio, $R$. 

The initial conditions are described by a shape and a perturbation in each of the fields, $T$, $S$, and $\nu_i$. 
To monitor the influence of the shape, we run three simulations of different initial boundary thickness ratios, $R_\textrm{init} \in \{10,4,2,1.5,1\}$. 
These simulations are initially perturbed in the salinity and temperature fields (see Table \ref{tab_simulations_sbm_3d_ic}) to quicken the transition to turbulence. 
After an initial transient all simulations tend towards a similar final boundary thickness ratio of just below two (see Figure \ref{fig_R_time})---independent of their shape. 
A similar behaviour was already observed before by \cite{carpenter_simulations_2012} for almost all of their simulations (see their Fig. 5). 

To monitor the influence of perturbation, we compare simulations that initially evolve diffusively to simulations that initially evolve turbulently. 
Those simulations with $R_\textrm{init} \leq 1.5$ result in an initially stably stratified water column across the whole domain. They initially evolve diffusively. 
Only after the temperature profile has diffused sufficiently (far into the domain where salinity does hardly vary anymore), does a buoyancy-reversal instability build up and promote convection. 
Those simulations with a perturbation in the velocity field initially evolve turbulently. 
After an initial transient all simulations tend towards a similar final boundary thickness ratio of just below two (see Figure \ref{fig_R_time})---independent of their perturbation. 

The observed boundary thickness ratio, $R$, is independent of the initial conditions. 
$R$ is thus intrinsic to the system that is wholly defined by $\{\textrm{Pr}, \textrm{Le}, R_\rho^s\}$. 
The flux ratio for $\textrm{Pr}=10$, $\textrm{Le}=4$ and $R_\rho^s=10$ targets about 1.8.

\subsection*{Bulk Heat Exchange Coefficient} 
\label{sec_bulk}
\begin{figure*}[tbp]
\begin{center}
  \includegraphics[width=.7\linewidth]{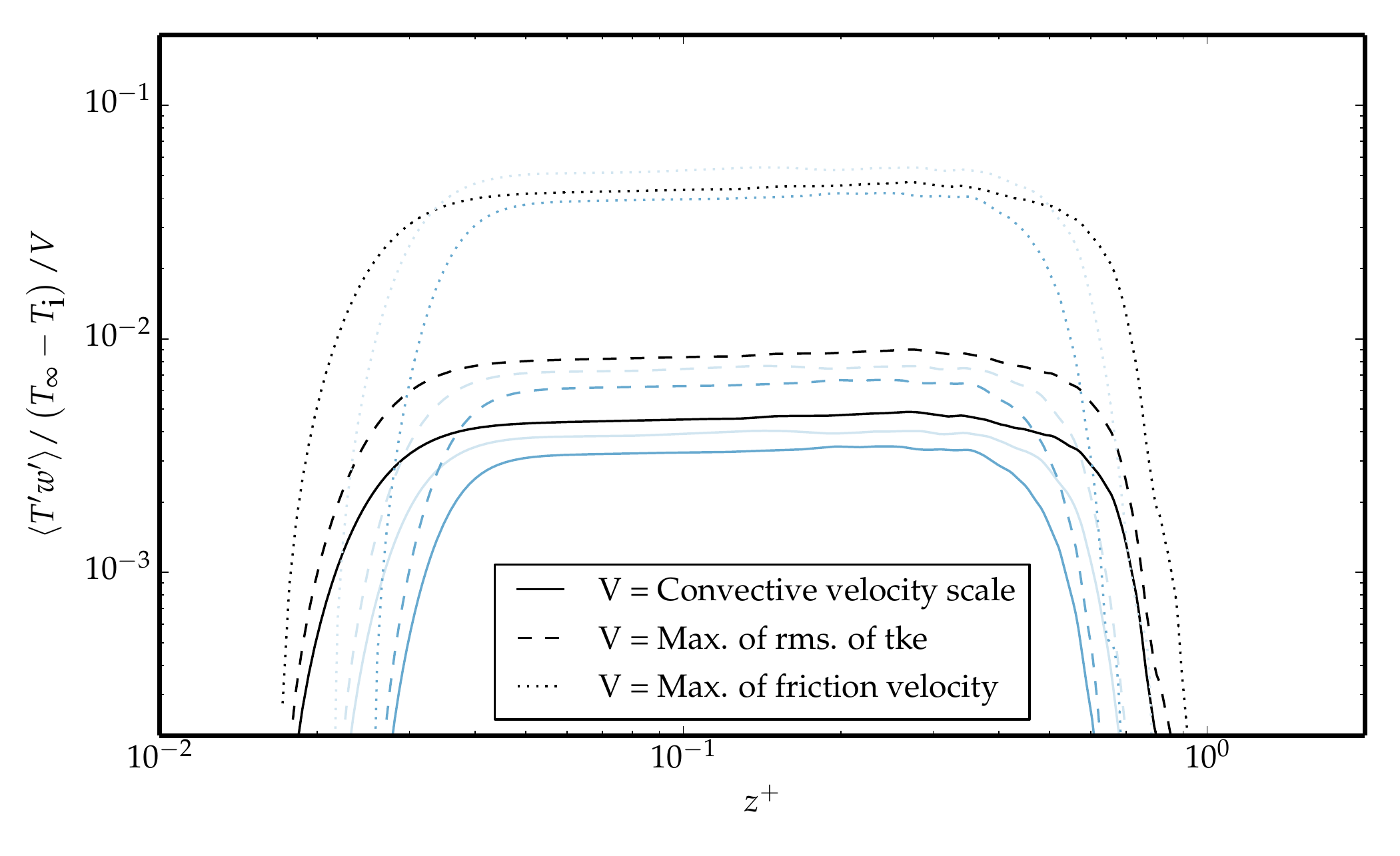}
\caption[]{\small{
Turbulent heat flux in units of the temperature scale $\left(T_\infty-T_\textrm{i}\right)$ times the velocity scale $V$. Convective velocity scale, $\nu_*$, (solid lines) maximum of the root mean square of the turbulent kinetic energy, $\textrm{max}(k_\textrm{rms})$, (dashed lines) and the friction velocity, $u_\textrm{*0}$, (dotted lines) are used as velocity scale, $V$. 
The convective velocity scale is generally used to compare to mean-shear velocities of forced convection setups. 
Colours indicate the density ratio of the simulation as given in Figure \ref{fig_rS1rS2_R}a. 
}}
\label{fig_alpha_y}
\end{center}
\end{figure*}

In free-convection setups, bulk scalar exchange coefficients are generally inferred from the turbulent flux in the outer layer that is scaled by the scalar scale and the convective velocity scale. 
For the simulations presented in this work, this yields bulk heat exchange coefficients of 0.003 to 0.0045 (see Figure \ref{fig_alpha_y}, solid lines). 

In field measurements, the friction velocity, $u_\textrm{*0}$ is used for the same purpose because one generally copes with mixed convection. 
For lack of a precise definition at which distance from the surface the friction velocity is to be measured, we estimate bulk heat exchange coefficients from our simulations by using the maximum friction velocities encountered across the domain (see Figure \ref{fig_alpha_y}, dotted lines). 

For the sake of completeness, we also provide the bulk heat exchange coefficients that can be estimated by using the maximum of the turbulent-kinetic energy rms.

\subsection*{Mixed-layer scalar fraction}
\label{sec_ml}
\begin{figure*}[tbp]
\begin{center}
  \includegraphics[width=.7\linewidth]{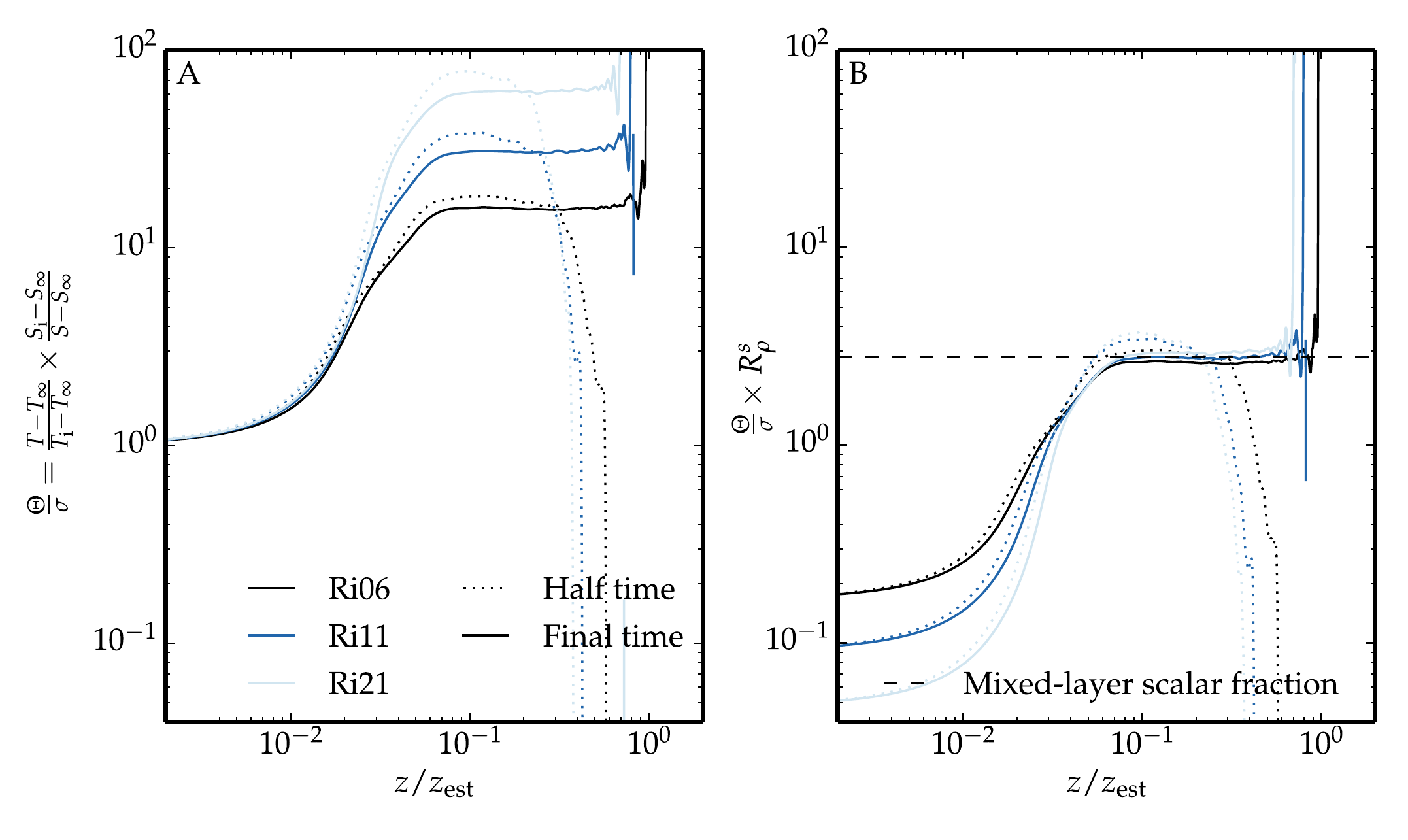}
\caption[]{\small{
a) Fraction between temperature scalar, $\theta$, and salinity scalar, $\sigma$, at final simulation time (solid lines) and half the simulation time (dotted lines) for different density ratios, $R_\rho^s$.
Colours indicate the density ratio of the simulation. 
b) Fraction between temperature scalar and salinity scalar scaled by the density ratio, $R_\rho^s$.
}}
\label{fig_rS1rS2_R}
\end{center}
\end{figure*}

The mixing in the outer layer (sometimes referred to as mixed layer) is mainly determined by turbulence. 
If the outer layer is well-mixed, temperature and salinity will be distributed homogeneously. 
In the following, we report our observation of how temperature is mixed compared to salinity. 

The mixing in the outer layer is driven by a buoyancy-reversal instability. 
This buoyancy-reversal instability is favoured by a temperature--salinity fraction that equals the density ratio, $R_\rho^s$:
\begin{linenomath}
\begin{align}
\label{eq_b_bilinear_nd}
&\frac{b}{b_m} = R_\rho^s~\sigma  - \theta \textrm{,}
\end{align}
\end{linenomath}
Temperature and salinity must be entrained in proportion $R_\rho^s$ to sustain mixing. 
In free convection, a homogeneously mixed outer layer should exhibit $\theta / \sigma \sim R_\rho^s$. 
Our simulations support this argument and provide the scaling, 
\begin{linenomath}
\begin{align}
\sigma_\textrm{ml} = \frac{1}{2.8} ~ \theta_\textrm{ml} ~ R_\rho^s \textrm{,}
\label{eq_ml_scalar}
\end{align}
\end{linenomath}
where $_\textrm{ml}$ refers to quantities in the outer layer (see Figure \ref{fig_rS1rS2_R}).

\section*{Acknowledgments}
Support from the Max Planck Society through its Max Planck Research Groups program
is gratefully acknowledged. 
Computational resources were supplied by J\"ulich Supercomputing Center. 
The work of \cite{ankit_rohatgi_version_2014} has been helpful in the course of this study. 

Primary data and scripts used in the analysis and other supplementary material
that may be useful in reproducing the author's work are archived by the Max Planck
Institute for Meteorology and can be obtained by contacting publications@mpimet.mpg.de

\bibliographystyle{copernicus}
\bibliography{Paper3-Latex}

\end{document}